# There's Plenty of Surface at the Bottom: Surface Chemistry Enhances Light Absorption by Colloidal Semiconductor Nanocrystals

*Carlo Giansante*

*NANOTEC-CNR Istituto di Nanotecnologia, via per Arnesano, 73100 Lecce, Italy*

*e-mail: carlo.giansante@nanotec.cnr.it

**Abstract.**

The paraphrasis of Feynman's renowned statement is used to highlight the inherent relevance of surfaces −and interfaces− at the nanoscale. Here, the marked impact of surface chemistry on the light absorption by colloidal inorganic semiconductor nanocrystals is demonstrated as general. Chemical species at the surface (ligands) of colloidal metal chalcogenide quantum dots (QDs) are shown to induce broadband absorption enhancement and band gap reduction. A comprehensive library of chalcogenol(ate) ligands is exploited to infer the role of surface chemistry on the QD optical absorption properties: ligand chalcogen binding atoms mainly determine band gap reduction, related to *n*p occupied orbital contribution to the valence band edge, and mediates broadband absorption enhancement, fostered by π-conjugation of the ligand pendant moiety, with further contribution from electron donor substituents. These findings point to a description of colloidal QDs that may conceive ligands as part of the overall QD electronic structure, beyond common models derived from analogies with core/shell heterostructures depicting ligands as mere perturbation to the core properties.

## INTRODUCTION.

Colloidal nanoscopic crystals of inorganic semiconductor materials in the quantum confinement regime (commonly referred to as quantum dots, QDs) show size-dependent optical and electronic properties that can be largely affected by surface chemical modification. Indeed, the widespread recourse to post-synthesis strategies to replace pristine bulky, electrically insulating chemical species at the QD surface (referred to as ligands) has led to disclose the relevant role exerted by surface species on the optoelectronic properties of the QDs. Such a ligand role is evident already at the ground state (and by naked eyes) with the common observation of optical band gap shift[1-4] and, more recently and for PbS QDs only, broadband optical absorption enhancement.[5-7] Most of these studies involved metal chalcogenide QDs and small organic molecules yielding marked QD first exciton bathochromic shift (up to several hundred meV)[8] and absorption enhancement (up to almost three times),[9] generally larger than that induced by inorganic ligands. Little consensus exists on univocal description for such phenomena and the present study aims at contributing to their explanation: here, colloidal CdS, CdSe, and PbS QDs are ligand-exchanged with chalcogenol(ate) species from a comprehensive library of ligands with tailored binding groups (the chalcogen atoms), bearing conjugated (or saturated) pendant moieties, eventually substituted with electronically- or sterically-active groups (as depicted in Scheme 1). This approach is here exploited to elucidate the role of surface chemistry modification on QD optical absorption properties. Comparison with descriptions derived from analogies with core/shell heterostructures conceiving ligands as perturbation to the QD local electric field or potential energy landscape is provided. It arises a description of colloidal QDs that implies ligands as part of the overall QD electronic structure.

## RESULTS AND DISCUSSION.

**A chalocogenol(ate) ligand library.** A widely used, simple ligand as p-methylbenzenethiol (ArSH; also used as p-methylbenzenethiolate triethylammonium ionic couple, $ArS^{-}$) is conceived as framework to introduce: (*i*) different binding groups (selenium as in selenophenol, PhSeH, and oxygen as in phenol, PhOH and its bidentate analog benzoic acid, PhCOOH, compared to the sulfur in thiophenol, PhSH); (*ii*) backbones with different degree of conjugation (the aliphatic chain of butanethiol, AlSH, and the benzyl moiety of phenylmethanethiol, BzSH, compared with the aromatic pendant group of ArSH or PhSH); (*iii*) different substituents that are electronically-active (such as for p-aminobenzenethiolate, D- ArS¯, and p-trifluoromethylbenzenethiolate, A-ArS¯) and introduce steric hindrance (as for o-dimethylbenzenethiolate, SE-ArS¯). These ligands are shown in Scheme 1.

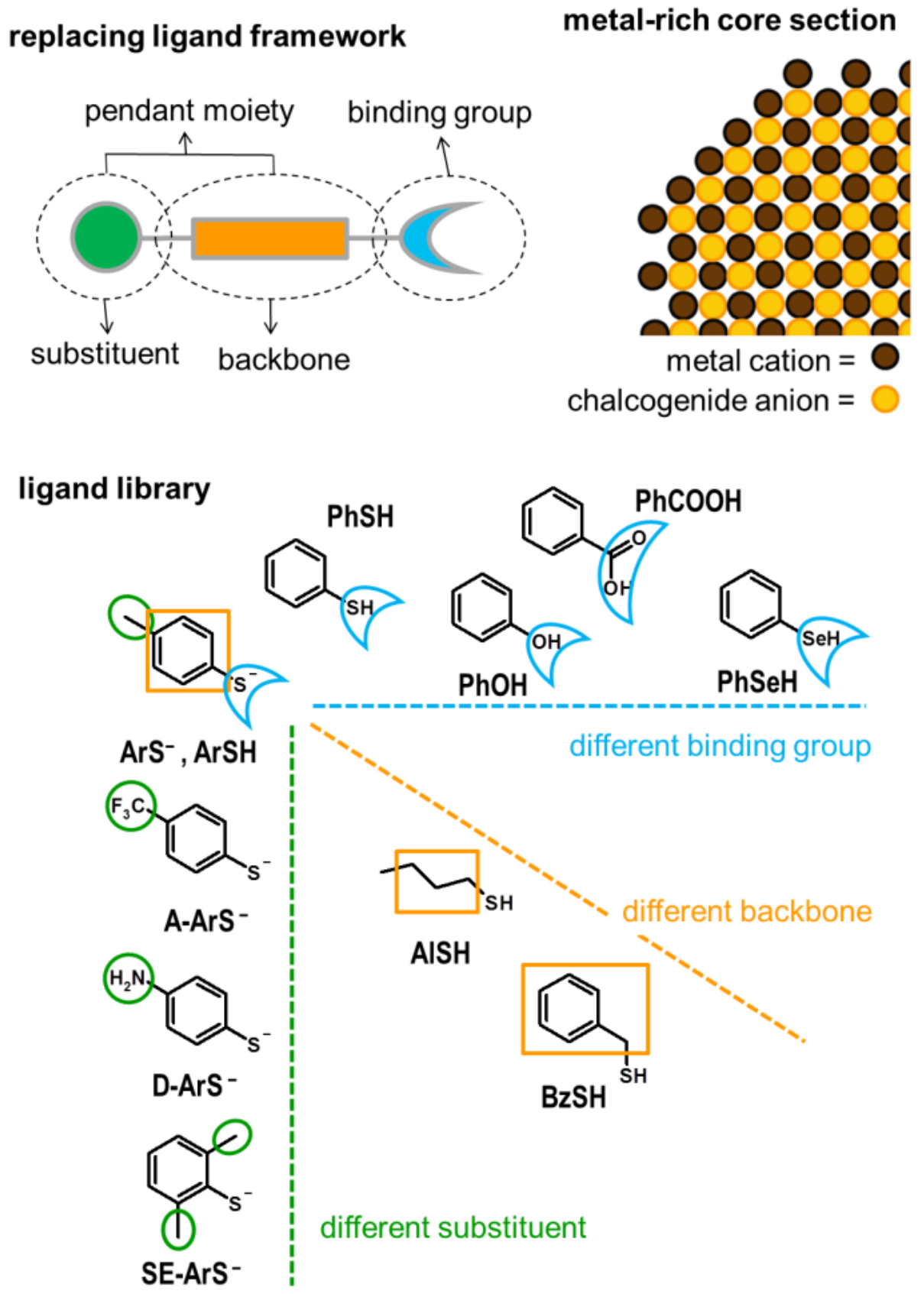

**Scheme 1.** Depiction of the replacing ligand framework at the QD surface conceived as constituted by a binding group and a pendant moiety, further consisting of a backbone bearing functional substituents. The ligand library used in this work is presented at the bottom of the figure with corresponding acronyms.

Ligand exchange is here performed in solution phase by simple addition of aliquots of chalcogenol ligand solutions to solutions of as-synthesized metal chalcogenide (CdS, CdSe, and PbS) QDs, thus permitting to neglect eventual inter-QD interactions due to excitonic/electronic coupling that may result upon chemically modifying the QD surface in solid phase and to control QD concentration, avoiding phase transfer or filtering that may hinder quantitative comparisons.

The metal-rich surface of colloidal metal chalcogenide QDs[16-18] has been repeatedly shown to have high affinity for chalcogenol molecules, displacing ligands coming from the synthetic procedure: the displacement can be nearly quantitative (as for carboxylates), particularly in solution-phase in which the QD surface is completely accessible. Indeed, equimolar amounts of $ArS^{-}$ are found as sufficient to quantitatively displace oleate ligands from the surface of QDs with cubic crystal structure and similar diameters of about 3 nm (as demonstrated by NMR spectra in Supplementary Figures S1-S3: upon $ArS^{-}$ addition, the vinylene peaks experience a slight shift and show fine structure indicative of unbound oleate).

**Ligand-induced changes in QD optical absorption.** The addition of $ArS^{-}$ to these QDs induces red shift of the first exciton peaks and increases optical absorption at all wavelengths (Figures 1a and 1b). Such spectral changes depend on $ArS^{-}$ concentration and do not show any light scattering ascribable to aggregation, with negligible extinction of the incident light at energies below the first excitonic peaks (Supplementary Figures S4-S6); spectral changes saturate at given $ArS^{-}$ to QD molar ratio beyond which the absorption spectra do not appreciably change, suggesting that the QD surface is no longer accessible to extra added ligands (Figures 1c and 1d) and permitting to estimate association constants (see Supplementary Information for details).[19] The first exciton peak bathochromic shift was assumed as the optical band gap reduction ($\Delta E_g$ in meV) and the broadband optical absorption enhancement was evaluated as the energy integrated absorption ratio ($\alpha/\alpha_0$, where $\alpha = \int A(\nu)d\nu$ integrated at energies below those reported to correspond to bulk-like values, i.e. coinciding with wavelengths of 340 nm for CdSe[20,21] and 400 nm for PbS[22], whereas 310 nm was arbitrarily assumed as bulk-like wavelength for CdS; $\alpha$ and $\alpha_0$ account for the energy integrated absorption of ligand-exchanged and as-synthesized QDs, respectively). For metal sulfide QDs upon $ArS^{-}$ addition, both the optical band gap reduction ($\Delta E_g$ of about 120 and 90 meV for CdS and PbS QDs, respectively) and the broadband absorption enhancement ($\alpha/\alpha_0$ of about 2.0 and 1.8 for CdS and PbS QDs, respectively) are marked and appreciably larger than those induced on CdSe QDs ($\Delta E_g$ of about 50 meV and $\alpha/\alpha_0$ of almost 1.5).

In order to broaden the significance of this experimental observation, $ArS^{-}$ was added also to wurzite Cd chalcogenide QDs: minor spectral changes are observed compared to those induced on zinc-blende Cd chalcogenide QDs (Figures S9 and S10; further discussion is in the Supplementary Information). As most ligands in the present library, and most ligands in the literature, are chalcogenol species rather than chalcogenolates, the effects of $ArS^{-}$ and ArSH were compared. In principle, neutral and anionic ligands (ArSH and $ArS^{-}$, respectively) can be expected to show different affinities for stoichiometric and metal-rich facets of the QDs; plots of the spectral changes induced by ArS¯ and ArSH on CdSe QDs do not suggest that eventually different Hammett constants and binding motifs of the ligands may markedly affect the $\alpha/\alpha_0$ and $\Delta E_g$ values (see Figures S11), in analogy to previous data on PbS QDs.[4] This permits the direct comparison of the effects exerted on QD optical absorption by all the ligands shown in Scheme 1.

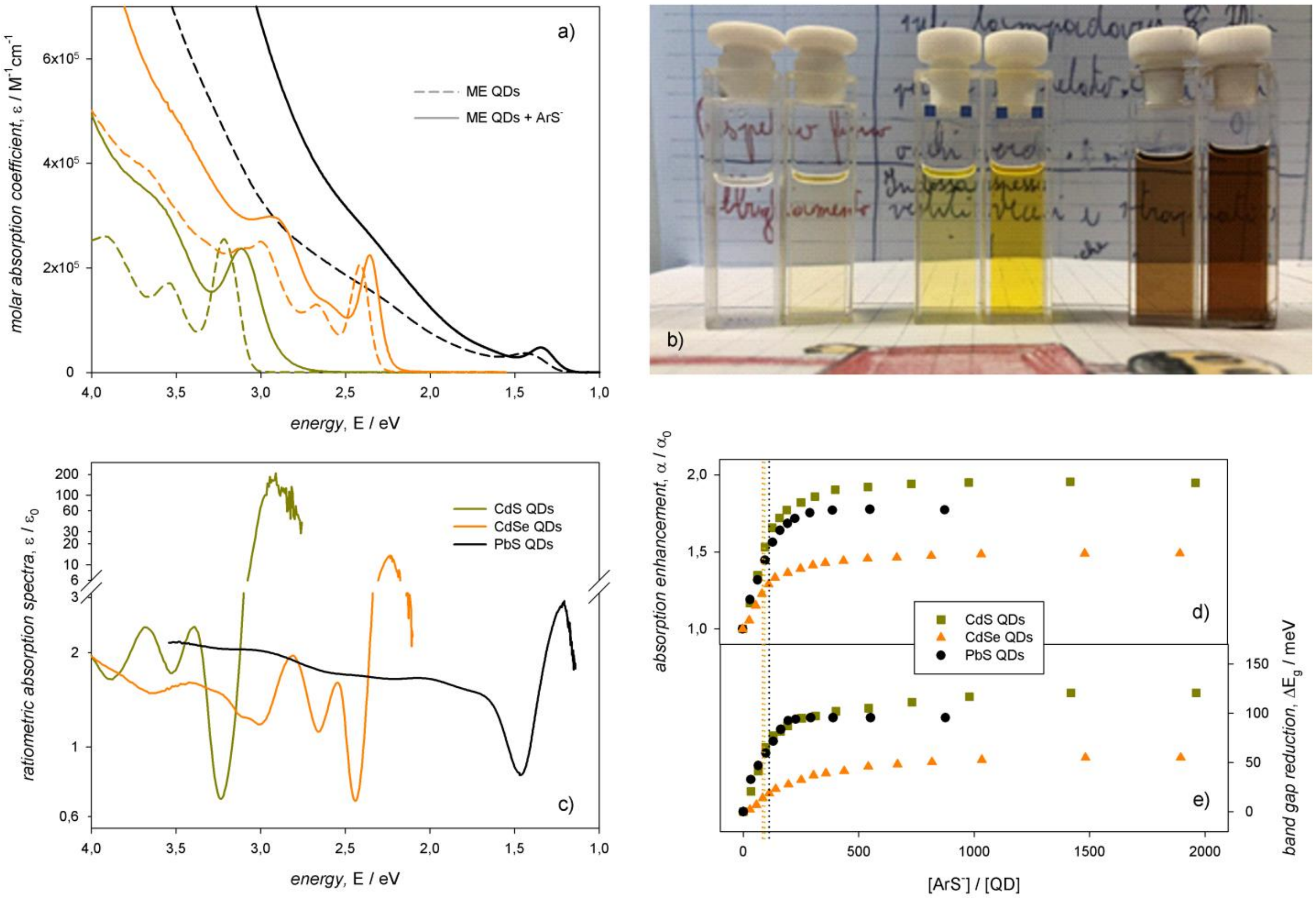

**Figure 1.** a) Molar absorption coefficients of as-synthesized CdS, CdSe, and PbS QDs with diameter of about 3 nm (dashed lines) and upon addition of ArS¯ (solid lines) up to plateau in dichloromethane solutions; b) daylight picture of such QD solutions. c) Ratiometric spectra of CdS, CdSe, and PbS QDs ligand exchanged with ArS¯ species. d) Plots of the ArS¯-induced QD broadband optical absorption enhancement ($\alpha/\alpha_0$, as the ratio of energy integrated absorption coefficients after and before ligand exchange). e) Plots of the ArS¯-induced QD optical band gap reduction ($\Delta E_g$, as the difference between first exciton peak energy before and after ligand exchange). Vertical dotted lines indicate the number of oleyl moieties per QD as determined by NMR.

**Ligand subunit role in QD optical absorption.** Complete ligand-dependence of QD optical absorption properties is presented in Figure 2, aimed at disentagling the role of each ligand subunit: namely, effects of the binding group and of the backbone and substituents that constitute the pendant moiety (panels a, b, and c of Figure 2, respectively) on CdS, CdSe, and PbS QDs (panels d, e, and f of Figure 2, respectively).

Ligand exchange on CdSe QDs with PhSH leads to very similar spectral changes to those induced by ArSH (and $ArS^-$; Figure S11). Replacement of the sulfur binding atom with oxygen, as for PhOH and its bidentate analog, PhCOOH, does not provide evidence for any appreciable spectral difference (Figures 2a and S12). The selenol analog, PhSeH, instead induced marked first exciton red shift and optical absorption enhancement ($\Delta E_g$ and $\alpha/\alpha_0$ of about 90 meV and 1.8, respectively), larger than those induced by PhSH (Figures 2a and S13), as already observed for diphenyldichalcogenides.[23] This suggests that the binding atoms enable QD spectral changes, which are larger when the chalcogen atoms of the replacing ligand coincide with the chalcogenide anions of the inorganic core.

The effect of ligand conjugation was evaluated comparing ArSH effect with that of benzyl, BzSH, and alkyl, AlSH, analogs (Figure 2b); the spectral changes induced by both BzSH and AlSH are similar and considerably smaller ($\Delta E_g$ and $\alpha/\alpha_0$ of about 20 meV and 1.1, respectively) than those induced by ArSH. This evidence confirms the pivotal importance of the ligand backbone conjugation in inducing the optical absorption enhancement; it also highlights a discrepancy with previous findings on PbS QDs, in which $\Delta E_g$ was found as independent on the ligand pendant moiety and instead mainly related to the ligand binding group.[4]

Upon introducing electronically-active substituents on the ligand conjugated backbone (as for D-$ArS^-$ and A-$ArS^-$), $\Delta E_g$ values similar to $ArS^-$ are observed, whereas larger $\alpha/\alpha_0$ values are found for D-$ArS^-$ only (Figure 2c). The presence of electron-donor substituents thus seems to contribute to further enhance optical absorption at all wavelengths,[9] although, upon using a library of substituted cinnamate ligands for PbS QDs, it has been suggested that is the band gap of the ligands themselves that determines the extent of the optical absorption enhancement.[24]

For metal sulfide (CdS and PbS) QDs, the addition of $ArS^-$, AlSH, and D-$ArS^-$ ligands induces roughly the same first exciton bathochromic shift, independently on the ligand pendant moiety (Figures 2d and 2f), whereas band gap reduction of CdSe QDs is lower for aliphatic AlSH compared to conjugated $ArS^-$ and D-$ArS^-$ (Figure 2e). For all metal chalcogenide QDs, the ligand exchange with D-$ArS^-$ induces optical absorption enhancement much larger than that induced by $ArS^-$, whereas much lower increase is observed for AlSH ligands.

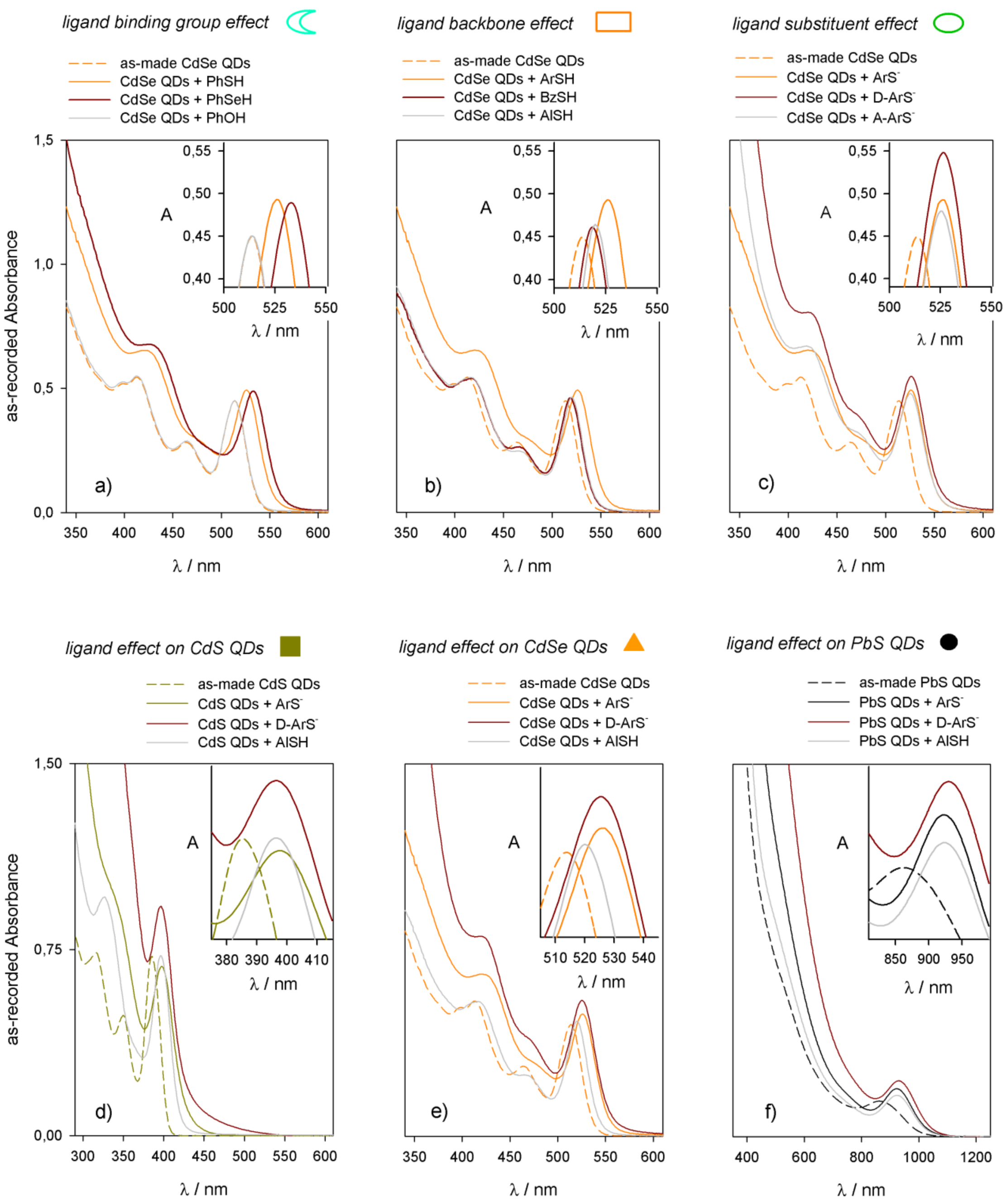

**Figure 2.** As-recorded absorption spectra of as-synthesized zinc-blende CdSe QDs (dashed lines) and upon addition of chalcogenol(ate) replacing ligands (solid lines) up to plateau, namely: a) PhOH, PhSH, PhSeH; b) ArSH, BzSH, AlSH; c) $ArS^{-}$, D-$ArS^{-}$, A-$ArS^{-}$. Insets show detail of the first exciton peak maxima. As-recorded absorption spectra of as-synthesized d) CdS, e) CdSe, and f) PbS) QDs with cubic crystal structure (dashed lines) upon addition of $ArS^{-}$, AlSH, D-$ArS^{-}$ (solid lines) up to plateau. Insets show detail of the first exciton peak maxima.

The size-dependent spectral changes observed for metal chalcogenide QDs upon ArS¯ addition are resumed in Figure 3: Figures 3a and 3b show the $\Delta E_g$ as function of the QD diameter and surface-to-volume ratio, respectively; Figures 3c and 3d present the optical absorption enhancement ($\mu/\mu_0$) at wavelengths with previously reported size-independent intrinsic absorption coefficients (set at 310 nm for CdS, 340 nm for CdSe and 400 nm for PbS). Both $\Delta E_g$ and $\mu/\mu_0$ are larger for smaller QDs (Figures 3a and 3c, respectively) with an almost linear dependence on QD surface-to-volume ratio (Figures 3b and 3d, respectively); the latter is however a preliminary conclusion as the size range here considered for Cd chalcogenide QDs is rather narrow and $\mu/\mu_0$ for PbS QDs may present a deviation from linearity for the smallest sample here evaluated.

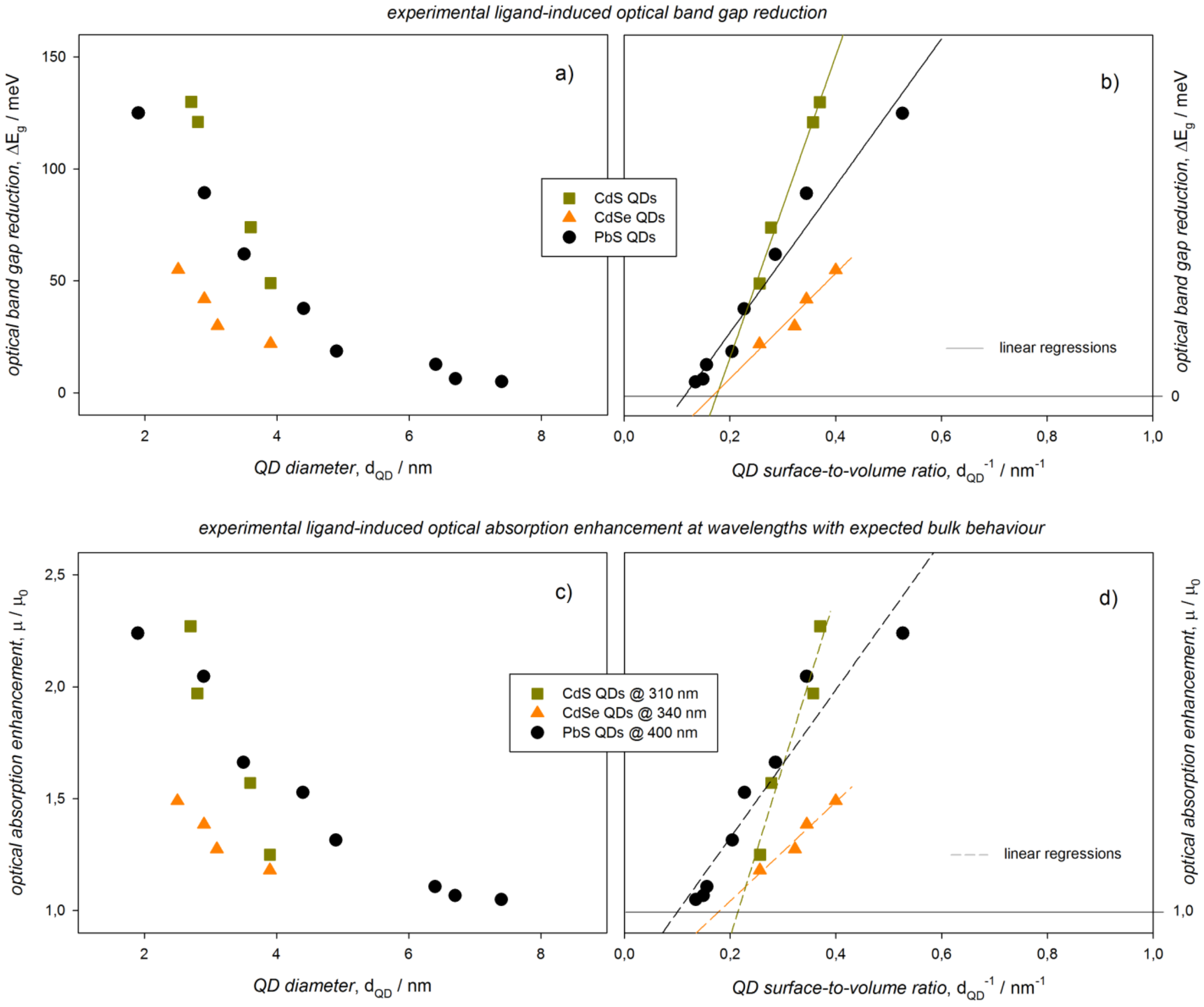

**Figure 3.** a,b) Plots of $\Delta E_g$ observed upon addition of ArS¯ to solutions of colloidal QDs (symbols) as function of (a) QD diameter and of (b) QD surface-to-volume ratio. c,d) Plots of $\mu/\mu_0$ observed upon addition of ArS¯ to solutions of colloidal QDs (symbols) as function of (c) QD diameter and of (d) QD surface-to-volume ratio.

**On the origins of ligand-induced QD spectral changes.** The above presented spectrophotometric data clearly confirm that the chalcogenol(ate) ligand effects on both $\Delta E_g$ and $\mu/\mu_0$ are general for metal chalcogenide QDs: the extents of the spectral changes depend on the inorganic core composition and size, showing an almost linear dependence on QD surface-to-volume ratio. The extents of both $\Delta E_g$ and $\alpha/\alpha_0$ (and $\mu/\mu_0$) also depend on the ligand themselves: the binding chalcogen atoms mainly determine the $\Delta E_g$ values as a result of the contribution of occupied *n*p orbitals to the valence band edge, which are instead poorly affected by the ligand pendant moiety (aliphatic and aromatic thiols induce similar $\Delta E_g$ values); this evidence comes with the noticeable exception of CdSe in which a difference between aliphatic and aromatic thiols is appreciable. In addition, larger $\Delta E_g$ values are obtained when binding atoms of ligands and anions of the core are the same chalcogen species. Aside from being the main responsible for optical band gap reduction, the ligand binding atoms also enable broadband optical absorption enhancement, but the extent of $\alpha/\alpha_0$ is mainly related to the $\pi$ conjugation of the ligand pendant moiety; it appears that the presence of electron-donor groups further enhances $\alpha/\alpha_0$.

These experimental evidences point to a description of colloidal QDs in which ligands do effectively take part to the overall electronic structure; the use of a comprehensive ligand library permits to infer the ligand role in the colloidal QD optical absorption and to heuristically define guidelines for predicting spectral changes upon ligand exchange. This description opposes to well-accepted models derived from analogies to core/shell heterostructures, in which ligands at the QD surface are regarded as perturbation of the electric field or the potential energy landscape at the core boundaries. Expected ligand-induced variations of both $\Delta E_g$ and $\mu/\mu_0$ upon considering ligands as dielectric shells or potential energy barriers are discussed below.

In analogy to core/shell heterostructures, the surface ligands had been indeed regarded as dielectric shell surrounding the inorganic core, at which dielectric mismatch may induce surface polarization affecting first exciton energy. Accordingly, the $\Delta E_g$ calculated upon exchanging oleate for benzenethiolate ligands is plotted in Figure 4a (details on calculations are given in the Supplementary Information),[25-28] showing values that are much smaller than those found experimentally (Figure 3a). In addition, it is also evident that the calculated ligand-induced dielectric confinement effect on the QD first exciton energy largely depends on the dielectric mismatch between the ligand shell and the inorganic core, yielding larger $\Delta E_g$ values for CdS than CdSe and PbS in agreement with their high-frequency[29] dielectric constants (5.7 for CdS,[25] for 10.2 CdSe,[30] and 17.2 for PbS[31]). Previous explanation for ligand-induced optical band gap reduction based on the description of ligands at the QD surface as dielectric shell indeed envisages anomalously large polarization effects.[3]

As further analogy to core/shell heterostructures, colloidal QDs had been described as hybrid heterojunctions, in which the height of the potential energy barrier at the ligand/core interface affects quantum confinement. Upon describing QD first exciton energy, within the effective mass approximation,[25] with the particle in a double spherical finite potential well model (details are given in the Supplementary Information),[32-34] the eigenfunction leakage on the benzenethiolate shell is appreciable for Cd chalcogenide QDs, whereas negligible for PbS QDs (Figure 4b) in contrast with the above presented experimental results. The calculated ligand-induced quantum confinement relaxation largely depends on the effective mass mismatch between the ligand shell (here considered as the rest mass of an electron, $m_e$) and the inorganic core, which yields eigenfunction leakage in the ligand shell larger for CdS than CdSe and negligible for PbS in agreement with their bulk hole effective masses ($0.80 \times m_e$ for CdS and $0.41 \times m_e$ for CdSe,[35] whereas of $0.085 \times m_e$ for PbS[36]). Previous description of ligand-induced QD optical band gap reduction assessing exciton delocalization based on the presumed alignment of ligand and core localized energy levels relies on the assumption that the effective mass of the organic ligand shell matches that of the inorganic core.[37]

At energies sufficiently above the band gap, quantum confinement effects are expected to vanish allowing the description of the linear optical absorption properties of QD dispersions within the framework of the Maxwell Garnett effective medium theory.[38-40] The optical absorption enhancement, $\mu/\mu_0$, calculated as the local field factor ratio upon exchanging oleate ligands for benzenethiolates (details are given in the Supplementary Information) is shown in Figure 4c and is larger than one order of magnitude than experimentally observed (Figure 3c). The calculated ligand-induced enhancement of the local field factor, which is directly proportional to the optical absorption enhancement, is larger for PbS QDs than Cd chalcogenide QDs; indeed, larger ligand/core dielectric mismatch implies larger dielectric confinement (Figure 4a) and therefore larger optical transition probabilities (Figure 4c). Previous description of QD optical absorption at energies far from the band gap based on the Maxwell Garnett effective medium theory neglects ligands[22] and may only include them as mere dielectrics, therefore being unable to predict the experimentally observed optical absorption enhancement.

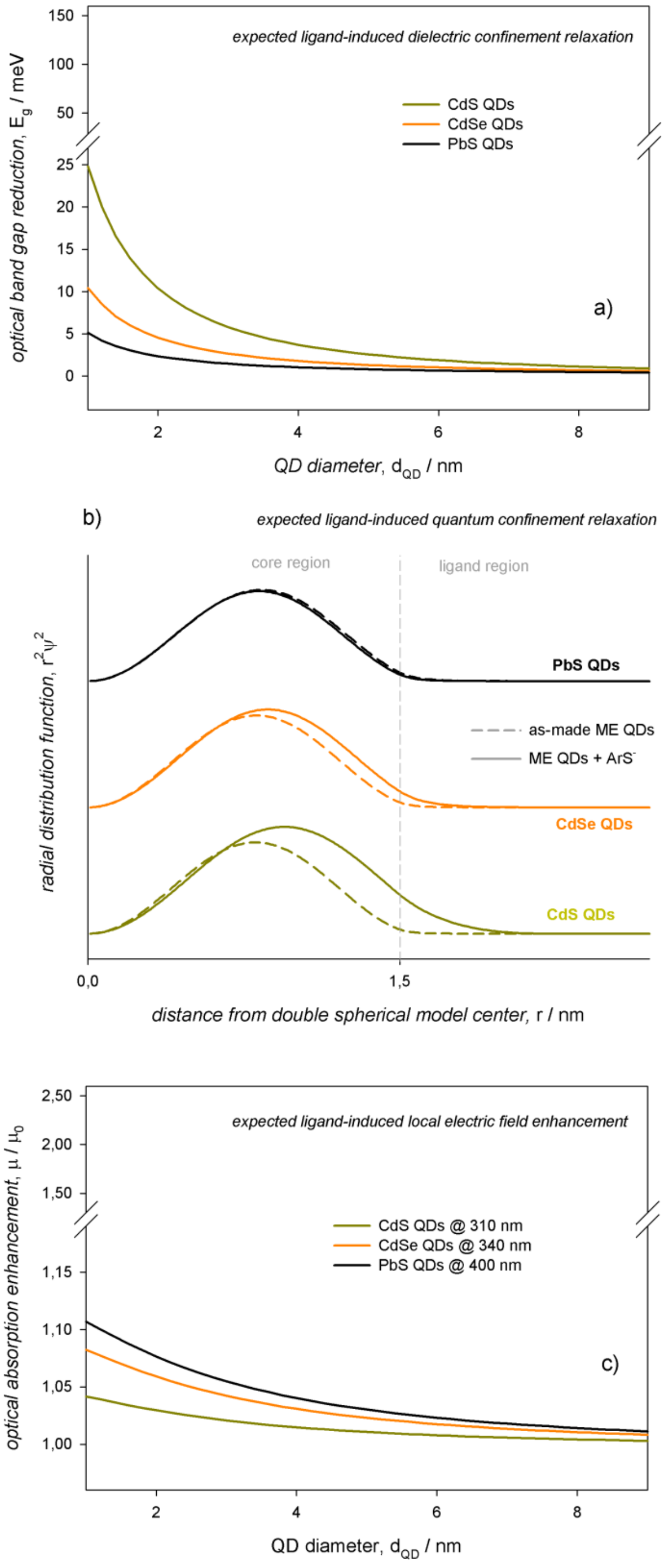

**Figure 4.** a) Optical band gap reduction predicted by polarization model for colloidal QDs ligand exchanged with $ArS^{-}$ (according to calculations reported in Supplementary equations S6-8). b) Radial distribution functions for QDs with 3.0 nm diameter (calculated according to Supplementary equations S9-14), with oleate and benzenethiolate ligands (dashed and solid lines, respectively). c) Optical absorption enhancement at wavelengths with expected bulk behavior for QDs ligand exchanged with $ArS^{-}$ (calculated according to Supplementary equations S15-18).

It is here shown that colloidal metal chalcogenide QDs undergo marked changes in their linear optical absorption properties upon post-synthesis surface chemistry modification with chalcogenol(ate) ligands. Broadband absorption enhancement and band gap reduction are experimentally observed and their origins are investigated by exploiting a comprehensive ligand library that allows to infer the ligand role in such spectral changes. Experiments clearly show that the binding (chalcogen) atoms of the replacing ligands play major role in the optical band gap reduction of colloidal QDs; the chalcogen *n*p orbitals of the replacing ligands are here suggested to contribute to the highest occupied states of the metal chalcogenide QDs, which are mainly constituted by *n*p chalcogenide orbitals. Chalcogen binding atoms also mediate broadband optical absorption enhancement that is fostered by conjugated ligand pendant moieties, further enhanced by electron-donor substituents; $\pi$ conjugation of the entire ligands is here envisaged to provide relevant contribution to the overall density of states.

These results point to a description of colloidal QDs in which ligands do take part to the overall QD electronic structure, beyond their common conception as localized states (e.g., forming, or healing, trap states). Albeit apparently simple, this conclusion could not have been reached according to core/shell analogies, which consider QDs as superposition of core and shell components and interfaces as abrupt. This implies that colloidal QDs constitute indecomposable chemical species and the description of their optical (and electronic) properties cannot neglect the potential relevance of the intrinsic ligand orbital contribution to the dominant core features.

**METHODS.**

*Materials.* All chemicals were of the highest purity available unless otherwise noted and were used as received; complete list appears in the Supplementary Information.

*Colloidal metal chalcogenide QD synthesis.* All QDs were synthesized in a three-neck flask connected to a standard Schlenk line setup under oxygen- and water-free conditions according to slightly modified recipes yielding: zinc-blende[10] and wurtzite[11] CdS QDs, zinc-blende[12] and wurtzite[11] CdSe QDs, and rock salt PbS QDs.[13] Briefly, colloidal QDs with cubic crystal lattice were synthesized upon hot-injection of the chalcogenide precursor in an octadecene solution of metal-oleate complex(es); size control was achieved by controlling oleic acid/metal oxide molar ratio. Colloidal QDs with hexagonal crystal lattice were synthesized upon swift injection of the chalcogenide precursors in a hot mixture of cadmium oxide, trioctylphosphine oxide, and octadecylphosphonic acid. Detailed synthetic and purification procedures appear in the Supplementary Information.

*Ligand Exchange Procedure.* Post-synthesis QD surface chemical modification was performed according to previous reports.[4-6,14,15] For spectrophotometric purposes, μL aliquots of mM solutions of the replacing ligands were added to μM solutions of the QDs in chlorinated solvents in a closed system (e.g., a vial or a quartz cuvette). Details appear in the Supplementary Information.

*Material characterization*. Spectrophotometric titration experiments were performed in quartz cuvettes with 1 cm path length and recorded with Varian Cary 5000 UV-Vis-NIR and Perkin Elmer Lambda 1050 spectrophotometers; $^1$H Nuclear Magnetic Resonance (NMR) spectra were recorded with a Bruker AV400 spectrometer operating at 400 MHz; Transmission Electron Microscopy (TEM) images were recorded with a Jeol Jem 1011 microscope operated at an accelerating voltage of 100 kV. Details are reported in the Supplementary Information.

**SUPPLEMENTARY INFORMATION.**

Details on the synthesis of colloidal QDs; ligand exchange procedure; further UV-Vis-NIR and NMR spectroscopic characterization; TEM images; theoretical calculations are available as Supplementary Information appearing below.

**REFERENCES.**

*Materials.*

All chemicals were of the highest purity available unless otherwise noted and were used as received. Lead oxide (PbO, 99.999%), Cadmium oxide (CdO, 99.5%), oleic acid (technical grade 90%), 1-octadecene (ODE, technical grade 90%), bis(trimethylsylil)sulfide (TMS, synthesis grade), p-methylbenzenethiol (ArSH, 98%), p-aminothiophenol (D-ArSH, 97%), p-(trifluoromethyl)thiophenol (A-ArSH, 97%), 2,6-dimethylthiophenol (SE-ArSH, 95%), 1-butanethiol (AlSH, 99%), benzyl mercaptane (BzSH, 99%), thiophenol (PhSH, 97%), benzeneselenol (PhSeH, 97%), phenol (PhOH), benzoic acid (PhCOOH), 3-methylbenzenethiol (95%) were purchased from Sigma-Aldrich. Tri-n-octylphosphine oxide (TOPO 99%), tri-n-octylphosphine (TOP, 97%), tri-n-butylphosphine (TBP, 99%), Sulfur (S, 99%), and Selenium (Se, 99,99%) were purchased from Strem Chemicals. Octadecylphosphonic acid (ODPA, 99%) was purchased from Polycarbon Industries. Triethylamine ($Et_3N$, ≥ 99.5%) was purchased from Fluka. All solvents were anhydrous and were used as received. Acetone (99.8%) was purchased from Merck. Acetonitrile (99.8%), dichloromethane (99.8%), hexane (95%), methanol (99.8%), tetrachloroethylene (99%), and toluene (99.8%) were purchased from Sigma-Aldrich.

*Colloidal metal chalcogenide QD synthesis.*

All QDs were synthesized in three-neck flasks connected to a standard Schlenk line setup under oxygen- and moisture-free conditions according to slightly modified, well-established recipes.

*Colloidal CdS QDs.*

In a typical synthesis yielding zinc-blende CdS QDs,[S1] 1 mmol of CdO (128 mg) was mixed with 3 mmol (850 mg) of oleic acid and 50 g of ODE. The mixture was vigorously stirred and deaerated through repeated cycles of vacuum application and purging with nitrogen at about 80 °C. Then, the mixture was heated to above 200 °C to allow dissolution of CdO until the solution became colorless and optically clear, indicating the formation of cadmium(II)-oleate complex(es). The solution was cooled at 80 °C and repeatedly subjected to vacuum in order to eventually remove water formed upon cadmium(II)-oleate complex formation. The solution was then heated again under nitrogen flow and stabilized at 300°C. At this point 0.5 mmol of S precursor (0.016g of S in 10g of ODE, previously prepared under nitrogen atmosphere) was swiftly injected and CdS QDs were allowed to grow at 250 °C for about 10 minutes. The heating mantle was then removed and the reaction quenched by compressed air. After the synthesis,

CdS QDs were transferred to a nitrogen-filled glove box. The QDs were repeatedly (three times, including the first step on crude product, are usually enough) precipitated using excess acetone (or methanol) and then redissolved in toluene, filtered through a 0.2 μm polytetrafluoroethylene membrane then stored at room temperature in a glove box for subsequent use as ~1 mM toluene solutions.

In a typical procedure yielding wurzite CdS QDs,[S2] TOPO (3.300g), ODPA (0.600g) and CdO (0.100g) were mixed in a three-neck flask, heated to ca. 90°C and repeatedly subjected to vacuum-nitrogen cycles. Under nitrogen atmosphere, the reaction mixture was then heated to above 300°C to dissolve the CdO until the solution turns optically clear and colorless. The heating mantle was removed to cool the flask to about 90 °C and again subjected to vacuum-nitrogen cycles. The temperature was then raised to 320°C and let stabilize then a mixture of TMS (0.170g) and TBP (3g) was swiftly injected. The nanocrystals were allowed to grow at 250°C to adjust their final diameter. The temperature was then quenched by compressed air. Afterwards, the QDs were precipitated with MeOH, washed by repeated re-dissolution in toluene and precipitation with the addition of MeOH, and finally kept in toluene.

*Colloidal CdSe QDs.*

In a typical synthesis yielding zinc-blende CdSe QDs,[S3] 0.3 mmol of CdO were mixed with 3 equivalents of oleic acid with 0.05 M concentration in ODE. The mixture was vigorously stirred and deaerated through repeated cycles of vacuum application and purging with nitrogen at about 80 °C. Then, the mixture was heated to above 200 °C to allow dissolution of CdO until the solution became colorless and optically clear, indicating the formation of cadmium(II)-oleate complex(es). The solution was cooled at 80 °C and repeatedly subjected to vacuum in order to eventually remove water formed upon cadmium(II)-oleate complex formation. The solution was then heated again under nitrogen flow and stabilized at 300°C. At this point 1 equivalent of Se powder dispersed, but not dissolved, in ODE was swiftly injected and CdSe QDs were allowed to grow at 260 °C. The heating mantle was then removed and the reaction quenched by compressed air. After the synthesis, CdSe QDs were transferred to a nitrogen-filled glove box. The QDs were repeatedly (three times, including the first step on crude product, are usually enough) precipitated using excess acetone (or methanol) and then redissolved in toluene, filtered through a 0.2 μm polytetrafluoroethylene membrane then stored at room temperature in a glove box for subsequent use as ~1 mM toluene solutions.

In a typical procedure yielding wurzite CdSe QDs,[S2] TOPO (3.000g), ODPA (0.280g) and CdO (0.060g) were mixed in a three-neck flask, heated to ca. 90°C and repeatedly subjected to vacuum-nitrogen cycles. Under nitrogen atmosphere, the reaction mixture was heated to above 300°C to dissolve the CdO until the solution turns optically clear and colorless. The heating mantle was removed to cool the flask to about 90 °C and again subjected to vacuum-nitrogen cycles. The temperature was then raised to the required injection temperature (around 380 °C) and let stabilize. Swift injection of the Se:TOP solution (0.058g Se + 0.360g TOP) followed. The injection temperature and the reaction time were modified in order to synthesize CdSe dots of different sizes. The reaction was then quenched by compressed air. Afterwards, the nanocrystals were precipitated with methanol, washed by repeated re-dissolution in toluene and precipitation with the addition of methanol, and finally kept in toluene.

*Colloidal PbS QDs.*

In a typical synthesis yielding rock salt PbS QDs,[S4] 2 mmol of PbO (450 mg) and 6 mmol (1700 mg) of oleic acid were mixed in 10 g of ODE. The mixture was vigorously stirred and deaerated through repeated cycles of vacuum application and purging with nitrogen at about 80 °C. The mixture was then heated to above 100 °C to allow dissolution of PbO until the solution became colorless and optically transparent, suggesting the complete formation of lead(II)-oleate complex(es). The solution was cooled at 80 °C and repeatedly subjected to vacuum in the attempt of removing water eventually released upon lead(II)-oleate complex formation. The solution was then heated again under nitrogen flow and the temperature stabilized at 110°C. At this point, 1 mmol of sulfur precursor (bis(trimethylsilyl)sulfide; 210 μL) in 2 mL of octadecene was swiftly injected. The heating mantle was immediately removed and the resulting colloidal solution was allowed to cool to room temperature. After the synthesis, PbS QDs were transferred to a nitrogen-filled glove box. The QDs were precipitated using excess acetone (1:4 vol/vol), centrifuged at 4000 rpm and then redissolved in toluene. Two additional precipitation-redissolution cycles were performed by using methanol and toluene. An estimate of solvent volumes (about 4 mL of toluene to dissolve the QD pellet and 2 mL of methanol to precipitate the QDs) is crucial to obtain subtly purified QDs, although inherently empirical. The PbS QD size was varied by changing the amount of oleic acid (from two to sixteen equivalents) added to PbO, keeping constant its total concentration in the reaction flask; eventual dilution yields larger nanocrystals. Oleylamine (2 mmol) was added to the Pb-oleate precursor to obtain PbS QDs with diameters below 2.5 nm. Stock toluene solutions with concentration of ~ 1 mM were prepared and stored at room temperature in the glove box for subsequent use.

*Replacing Ligand and QD Solution Proparation.*

Replacing ligand solutions were prepared in glove box by diluting as received chalcogenols or by adding one equivalent of triethylamine to dichloromethane 10-50 mM solutions of the thiols shown in Scheme 1: according to reported $pK_a$ (in water, 6.5 and 10.8 for ArSH,[S5] and $Et_3N{\cdot}H^+$,[S6] respectively), the equilibrium, though established in much less polar dichloromethane, shifts towards the formation of the ammonium thiolate ionic couple. A qualitative demonstration relies in the formation of needle crystals upon dichloromethane evaporation from solutions containing equimolar amounts of 3-methylbenzenethiol and triethylamine, which are both liquids at room temperature; empirically, thiol fetor is somehow reduced upon amine addition. Triethylamine was not added to AlSH that is much weaker acid than aromatic analogs ($pK_a$ for AlSH is 11.5);[S5] indeed, it has been previously shown that titration of PbS QDs with AlSH leads to spectral changes that are negligibly affected by the presence of triethylamine.[S7]

Colloidal QD solutions were prepared in glove by diluting as-synthesized, purified batches to the concentration of 0.2 - 10 μM (depending on QD size), which was spectrophotometrically determined by using the Lambert-Beer law and well-established relationship between QD size and molar absorption coefficients of colloidal CdS,[S8] CdSe,[S9] and PbS QDs.[S10] Dichloromethane was used as solvent for Cd chalcogenide QDs and for PbS QDs of diameter below 3 nm, whereas tetrachloroethylene was used for larger PbS QDs to avoid solvent absorption in the NIR spectral region; no appreciable differences were observed by using these two solvents for PbS QDs with diameter of about 3 nm.[S7] This also supports the negligible role played by dielectric effects on the optical band gap of colloidal QDs.

*Ligand Exchange Procedure.*

Post-synthesis QD surface chemical modification was performed in a closed system (e.g., a vial or a quartz cuvette as for spectrophotometric titration experiments) by adding aliquots of solutions of the replacing ligands (as described above) to dichloromethane (or tetrachloroethylene) solutions of colloidal QDs. Closed systems account for sealed containers that guarantee no exchange of matter with the surroundings: it specifically refers to the precise control of QD concentration, which is instead hampered in ligand exchange procedures that imply filtering[S11] or phase transfer.[S12]

The ligand exchange process suddenly occurs upon replacing ligand addition to QD solutions: this can be evaluated by naked eyes as band gap reduction for small CdS QDs shifts the optical absorption in the visible spectral range, thus turning the solution from colorless to yellow,[S13] whereas small PbS QDs undergo concomitant broadband optical absorption enhancement that makes the solution significantly darker.[S14] The modified absorption spectra are constant with time and do not show any light scattering ascribable to aggregation, as demonstrated by the negligible extinction of the incident light at energies below the first excitonic peak.

Swift addition of high ligand loads without shaking QD solutions cause precipitation. Do ligand exchange slowly.

Chalcogenol ligands stink! Use under fume hood; pour remnant solutions into bleach to reduce stench by oxidation; leave glassware in a bleach bath before washing (or disposal).

*UV-Vis-NIR Absorption Spectroscopy.*

The optical absorption spectra of colloidal QDs were measured in suprasil quartz cuvettes with 1 cm path length and were recorded with a Varian Cary 5000 UV-Vis-NIR spectrophotometer. The ligand-induced optical band-gap reduction reaches plateau at given ligand-to-QD molar ratios, beyond which the absorption spectrum does not appreciably change, suggesting that the QD surface is no longer accessible to extra added ligands and thus suggesting also quantitative ligand exchange (see Figure 2; this is valid for oleate-capped metal chalcogenide QDs, while plateau is not reached with phosphonate-capped metal chalcogenide QDs). [S7,S10,S13-S17]

*Transimission Electron Microscopy (TEM).*

TEM images were recorded with a Jeol Jem 1011 microscope operated at an accelerating voltage of 100 kV. Samples for analysis were prepared by dropping a QD solution onto carbon-coated Cu grids and then allowing the solvent to evaporate in a vapor controlled environment. QD diameters were determined by statistical analysis of TEM images of several hundreds of QDs with the ImageJ software. [S7,S10,S13-S17]

*Nuclear Magnetic Resonance Spectroscopy (NMR).*

NMR spectra were recorded with a Bruker AV400 spectrometer operating at 400 MHz on 0.1 mM solutions of ME QDs in $CDCl_3$ ($\delta$ = 7.26 ppm) at 293 K, in presence of ferrocene ($\delta$ = 4.17 ppm) as internal standard to determine oleate ligand content. The addition of thiolate ligands was performed in a vial to allow proper mixing with QDs. Upon $ArS^-$ addition, the vinylene peak ($\delta \approx 5.3$ ppm) slightly shifts and shows fine structure indicative of oleate displacement; concomitantly, the resonances of protons in orto and meta positions of the $ArS^{-}$ ligands ($\delta \approx 7.0$ and 7.2 ppm) appear, as shown in Figures S1-3.

*Theoretical Calculations.*

Thermodynamics of ligand binding at the QD surface (eqs. S1-5), dielectric and quantum confinement effects on QD first exciton energy (S6-S8 and S9-S14, respectively), and Maxwell Garnett effective medium theory applied to intrinsic absorption coefficients at energies far from the QD band gap (eqs. S15-18) were estimated by using Mathematica and Sigmaplot softwares.

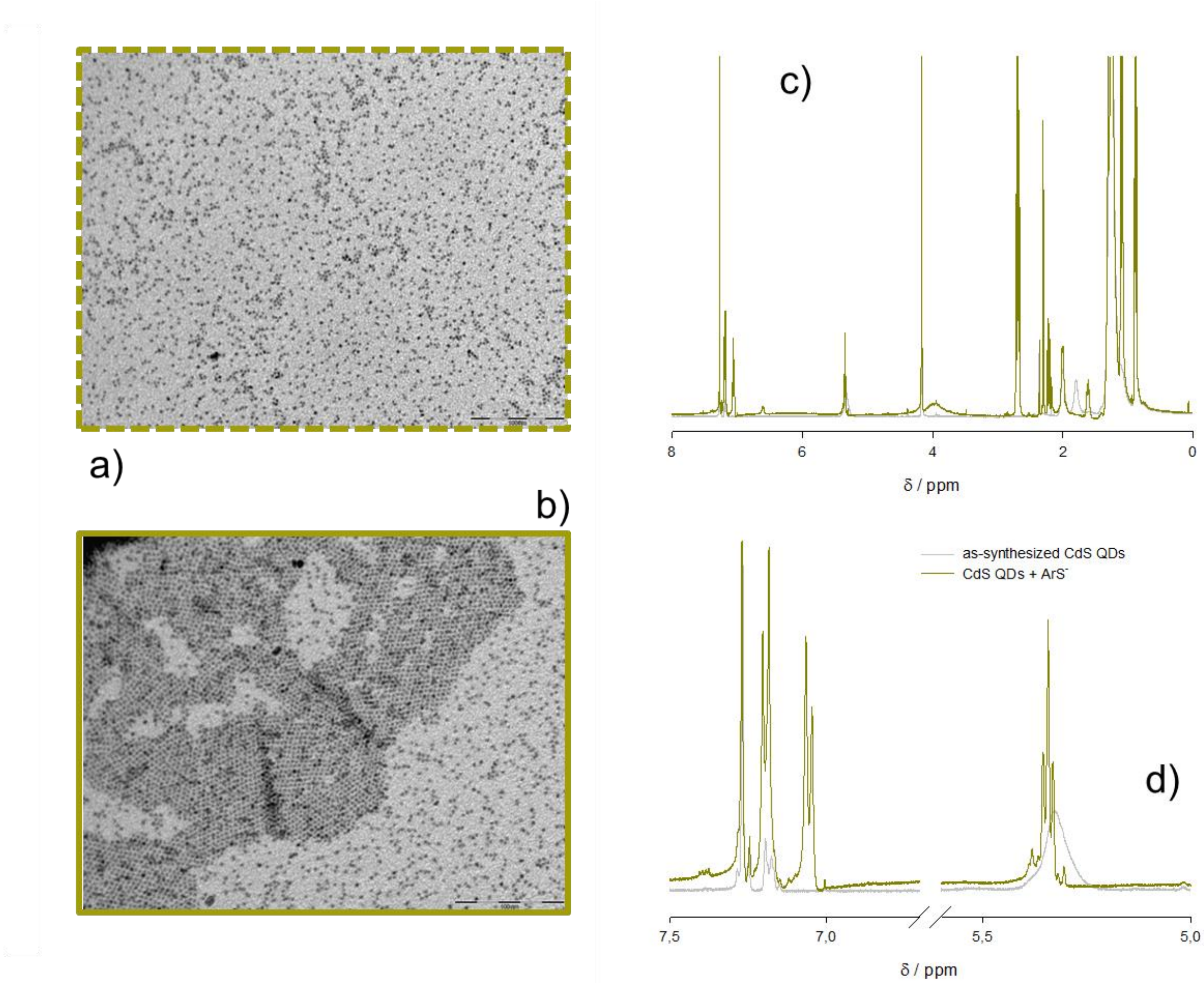

**Figure S1.** Transmission electron microscopy images of as-synthesized CdS QDs with diameter of about 2.8 nm (a) and upon addition of triethylammonium p-methylbenzenethiolate (b). NMR spectra (c) of as-synthesized CdS QDs (grey line) and upon addition of p-methylbenzenethiolate ligands (yellow line); panel (d) shows detail of the aromatic region.

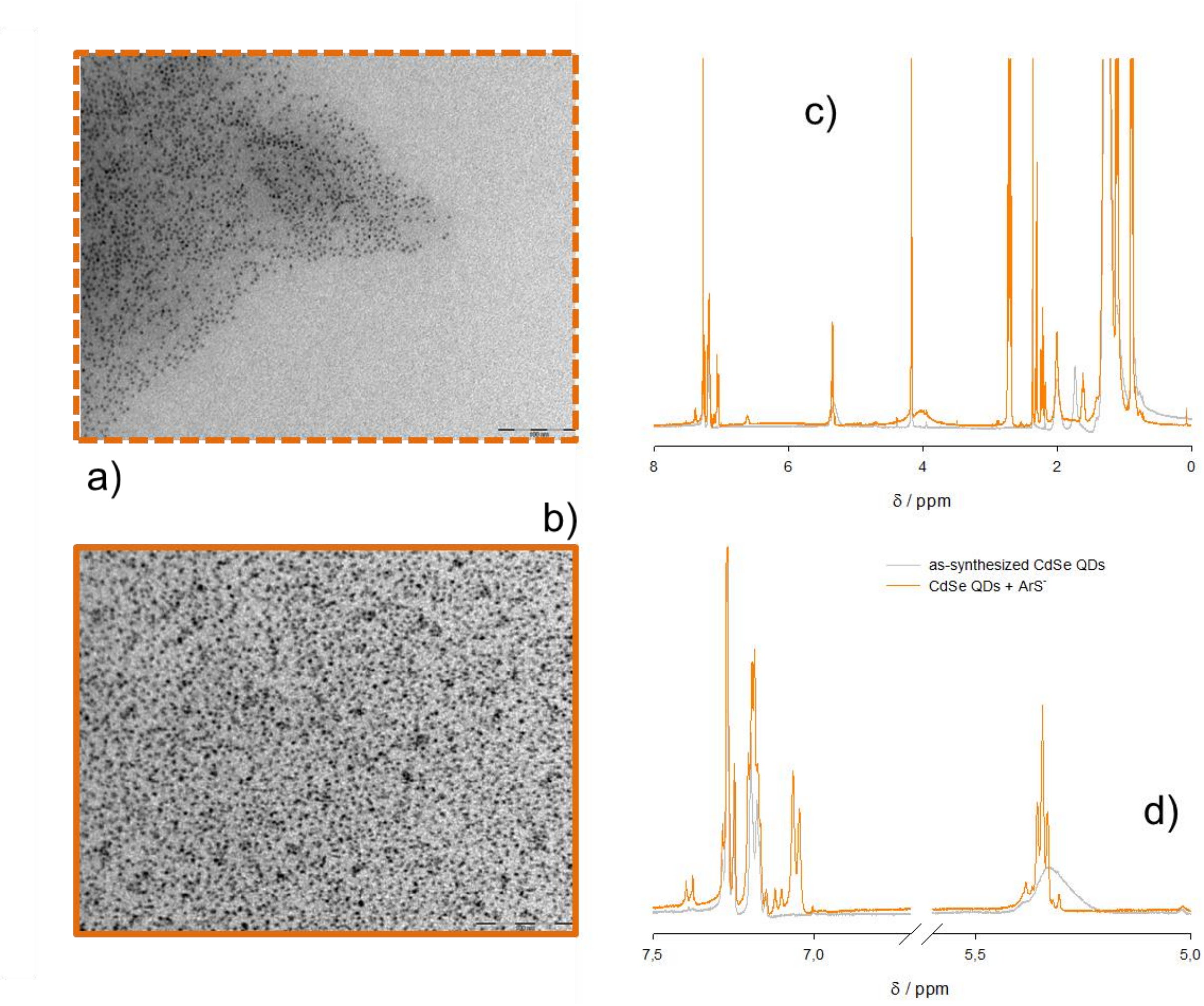

**Figure S2.** Transmission electron microscopy images of as-synthesized CdSe QDs with diameter of about 2.7 nm (a) and upon addition of triethylammonium p-methylbenzenethiolate (b). NMR spectra (c) of as-synthesized CdSe QDs (grey line) and upon addition of p-methylbenzenethiolate ligands (orange line); panel (d) shows detail of the aromatic region.

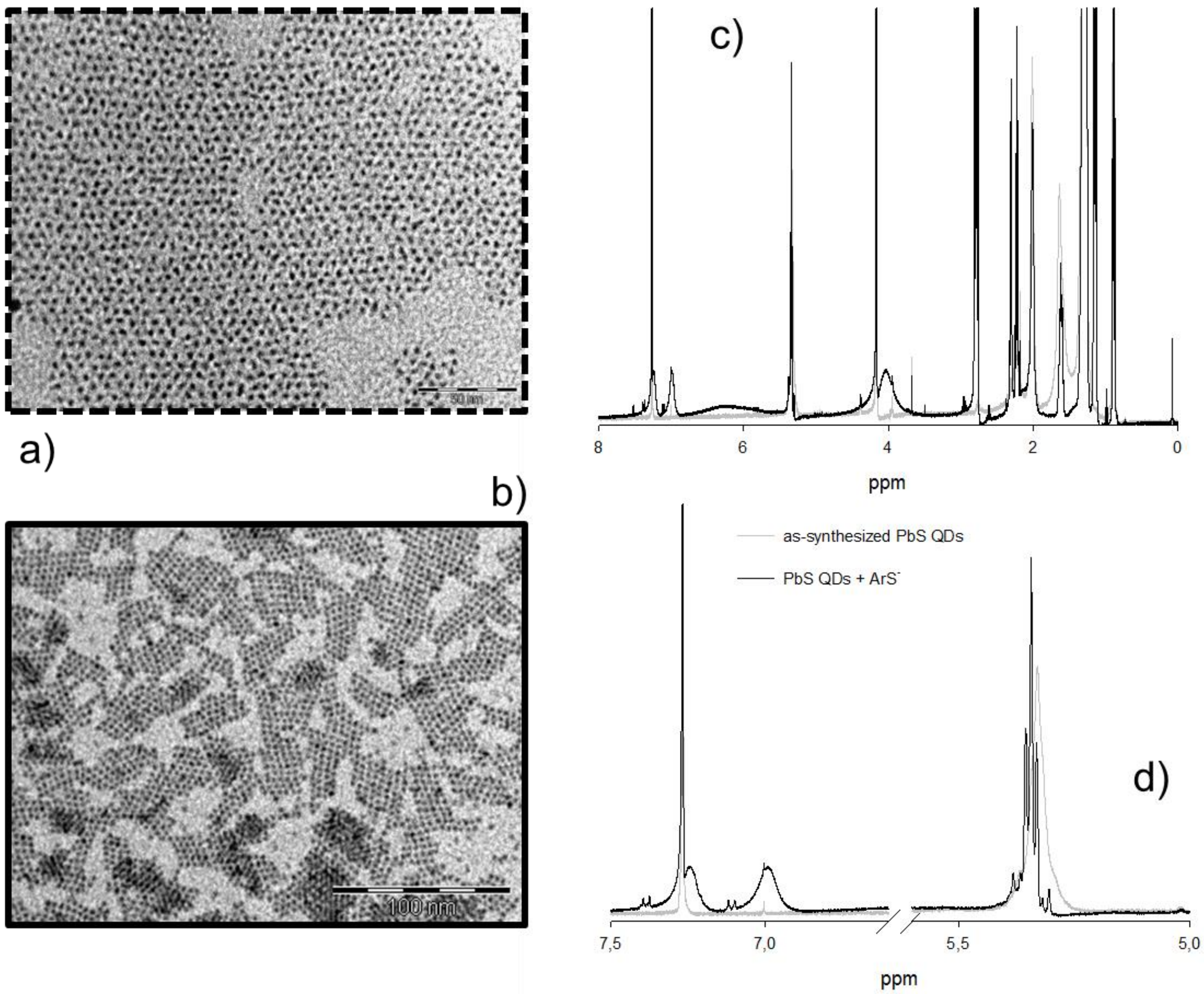

**Figure S3.** Transmission electron microscopy images of as-synthesized PbS QDs with diameter of about 2.9 nm (a) and upon addition of triethylammonium p-methylbenzenethiolate (b). NMR spectra (c) of as-synthesized PbS QDs (grey line) and upon addition of p-methylbenzenethiolate ligands (black line); panel (d) shows detail of the aromatic region.

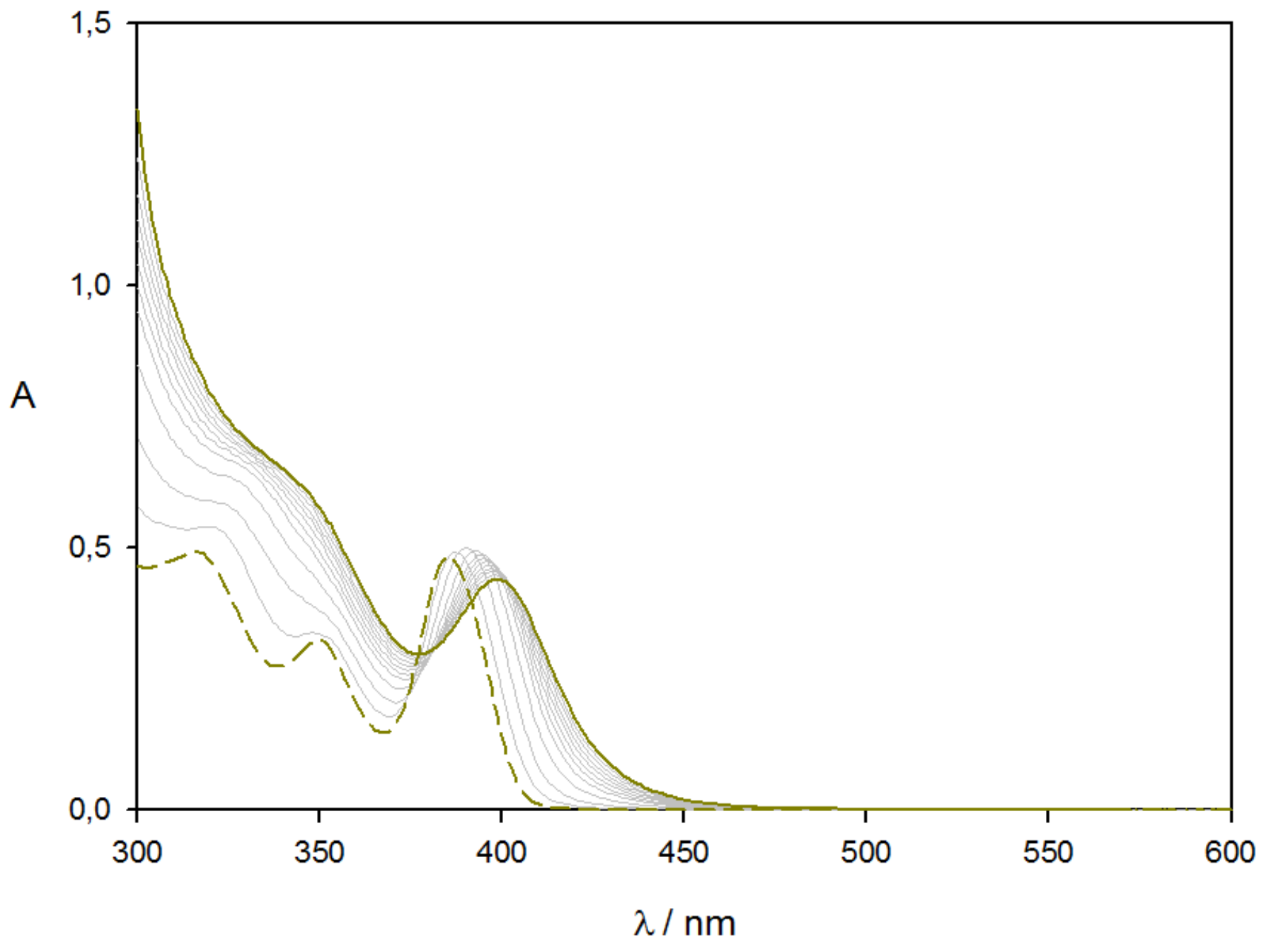

**Figure S4.** Optical absorption spectra of CdS QDs with diameter of about 2.8 nm upon titration with freshly prepared p-methylbenzenethiolate ligand solutions.

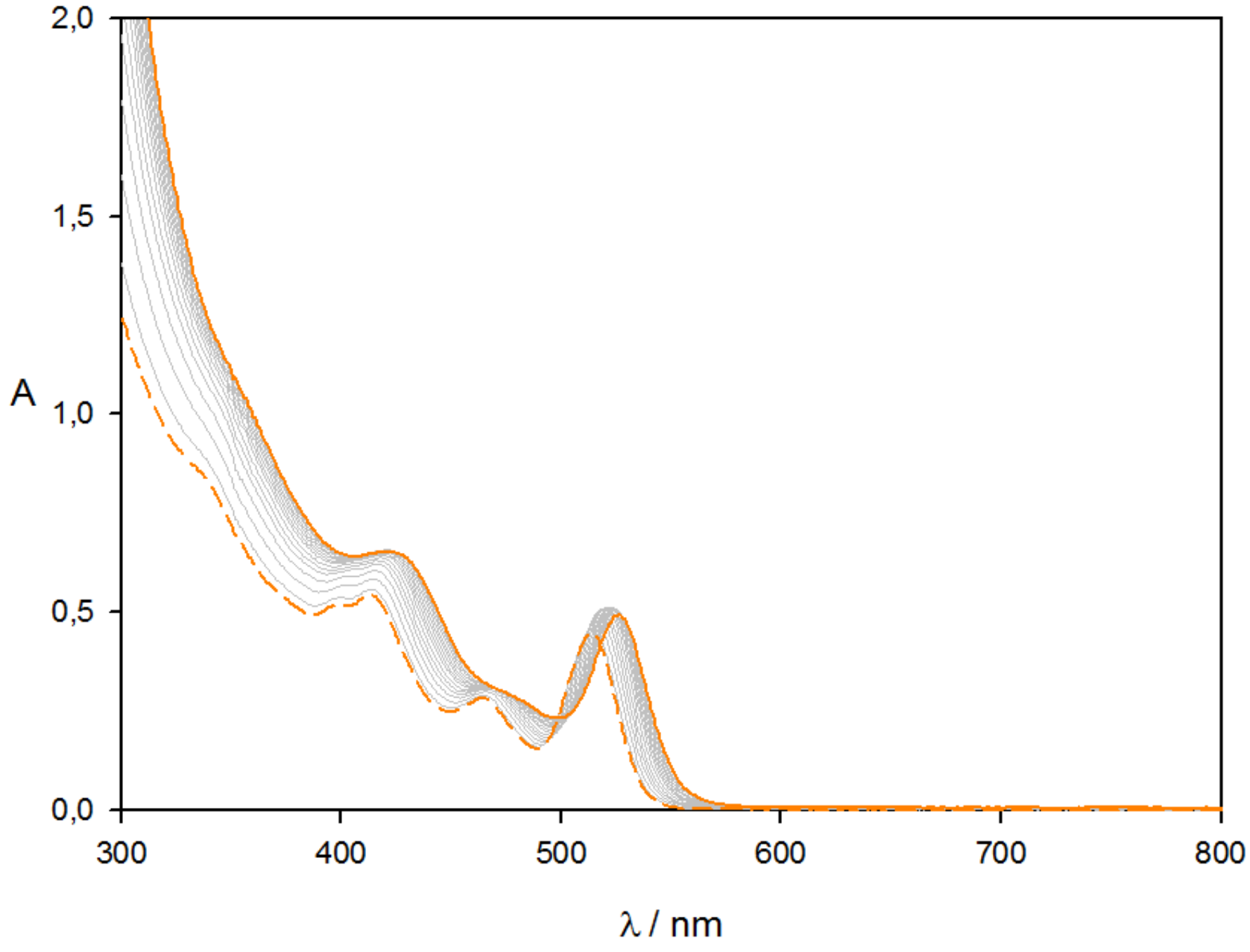

**Figure S5.** Optical absorption spectra of CdSe QDs with diameter of about 2.7 nm upon titration with freshly prepared p-methylbenzenethiolate ligand solutions.

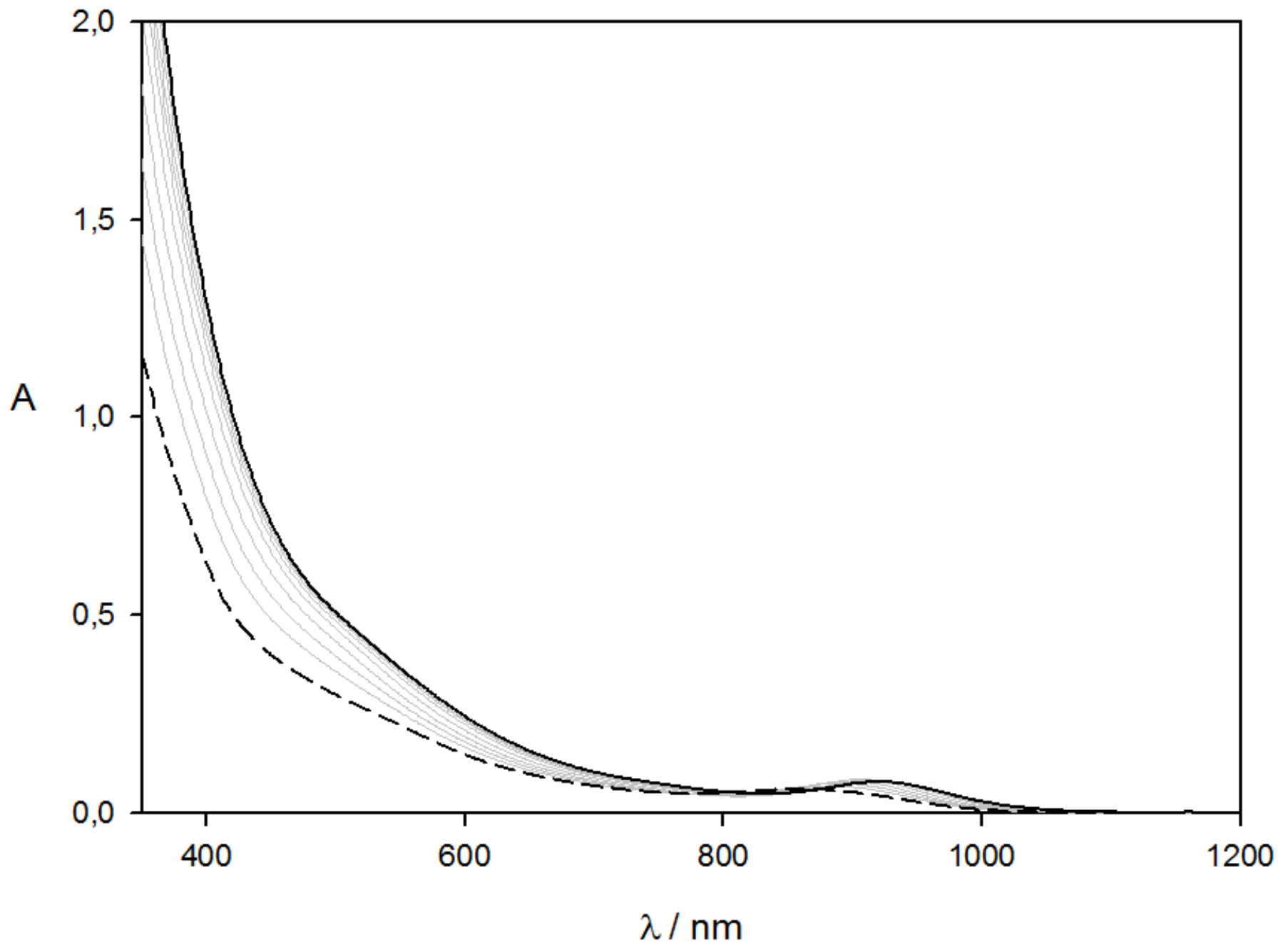

**Figure S6.** Optical absorption spectra of PbS QDs with diameter of about 2.9 nm upon titration with freshly prepared p-methylbenzenethiolate ligand solutions.

Upon assuming that the plateau indicates complete ligand exchange and that the number of binding sites per QD does coincide with the number of oleyl moieties (vertical dotted lines in Figures 1c and 1d), it is possible to estimate the association constants for ArS‾ binding at the QD surface: namely, $0.95\times10^4$ $M^{-1}$ for CdS QDs, $0.72\times10^4$ $M^{-1}$ for CdSe QDs, and $1.8\times10^4$ $M^{-1}$ for PbS QDs (see Supplementary Figure S7). The steeper rise and more flat plateau of PbS QDs compared to Cd chalcogenide QDs qualitatively suggest tighter binding of ArS‾ to excess Pb atoms compared to Cd binding sites at the QD surface (see normalized data as Supplementary Figure S8); as equimolar amounts of ArS‾ seem sufficient to quantitatively displace oleate ligands from the QD surface according to NMR spectra (Supplementary Figure S1-S3), whereas the number of ArS‾ ligands necessary to reach plateau in spectrophotometric measurements exceeds the number of bound oleyl moieties per QD, the displacement of species other than oleate (such as hydroxyl moieties, already shown to coordinate excess Pb atoms at the PbS QD surface – commonly, one oleyl per excess Pb atom was found in Pb chalcogenide QDs–[S16,S19,S20]) may be plausible, although this issue is not further investigated in this work.

Given the equilibrium reaction between the as-synthesized ME QDs and the added ligands (ArS) leading to ME/ArS QDs:

$$ME + ArS \leftrightarrow ME / ArS \qquad \text{(S1)}$$

the association constant of the reaction is

$$K_{ass} = \frac{[ME/ArS]_{eq}}{[ME]_{eq}[ArS]_{eq}} = \frac{[ME/ArS]_{eq}}{\left([ME]_0 - [ME/ArS]_{eq}\right)\left([ArS]_0 - [ME/ArS]_{eq}\right)} \qquad \text{(S2)}$$

The observed absorbance at wavelengths sufficiently far from the band gap (λ) upon ArS‾ addition, is related to the equilibrium concentration of the ME/ArS QDs formed in solution,

$$A_{obs}{}^{\lambda} = \varepsilon_{ME}{}^{\lambda}\left([ME]_0 - [ME/ArS]_{eq}\right) + \varepsilon_{ME/ArS}{}^{\lambda}\left([ME/ArS]_{eq}\right) + \varepsilon_{ArS}{}^{\lambda}\left([ArS]_0 - [ME/ArS]_{eq}\right) \quad \text{(S3)}$$

where $\varepsilon_{ME}{}^{\lambda}$ is as in references S8-10 and $\varepsilon_{ArS}{}^{\lambda} = 0$, thus the equilibrium concentration of the ME/ArS QDs can be derived by eq. S1:

$$[ME/ArS]_{eq} = \frac{1}{2K_{ass}}\left(a - \sqrt{a^2 - 4K_{ass}{}^2[ME]_0[ArS]_0}\right) \quad \text{(S4)}$$

where,

$$a = K_{ass}[ME]_0 + K_{ass}[ArS]_0 + 1 \quad \text{(S5)}$$

Using this set of equations and considering the number of $ArS^{-}$ ligands per oleyl moieties per QD as determined by quantitative NMR, fitting of the experimental absorbance values as in Figure S7 gives the binding constant values reported in the main text ($0.95\times10^4$ $M^{-1}$ for CdS QDs, $0.72\times10^4$ $M^{-1}$ for CdSe QDs, and $1.8\times10^4$ $M^{-1}$ for PbS QDs).[S18]

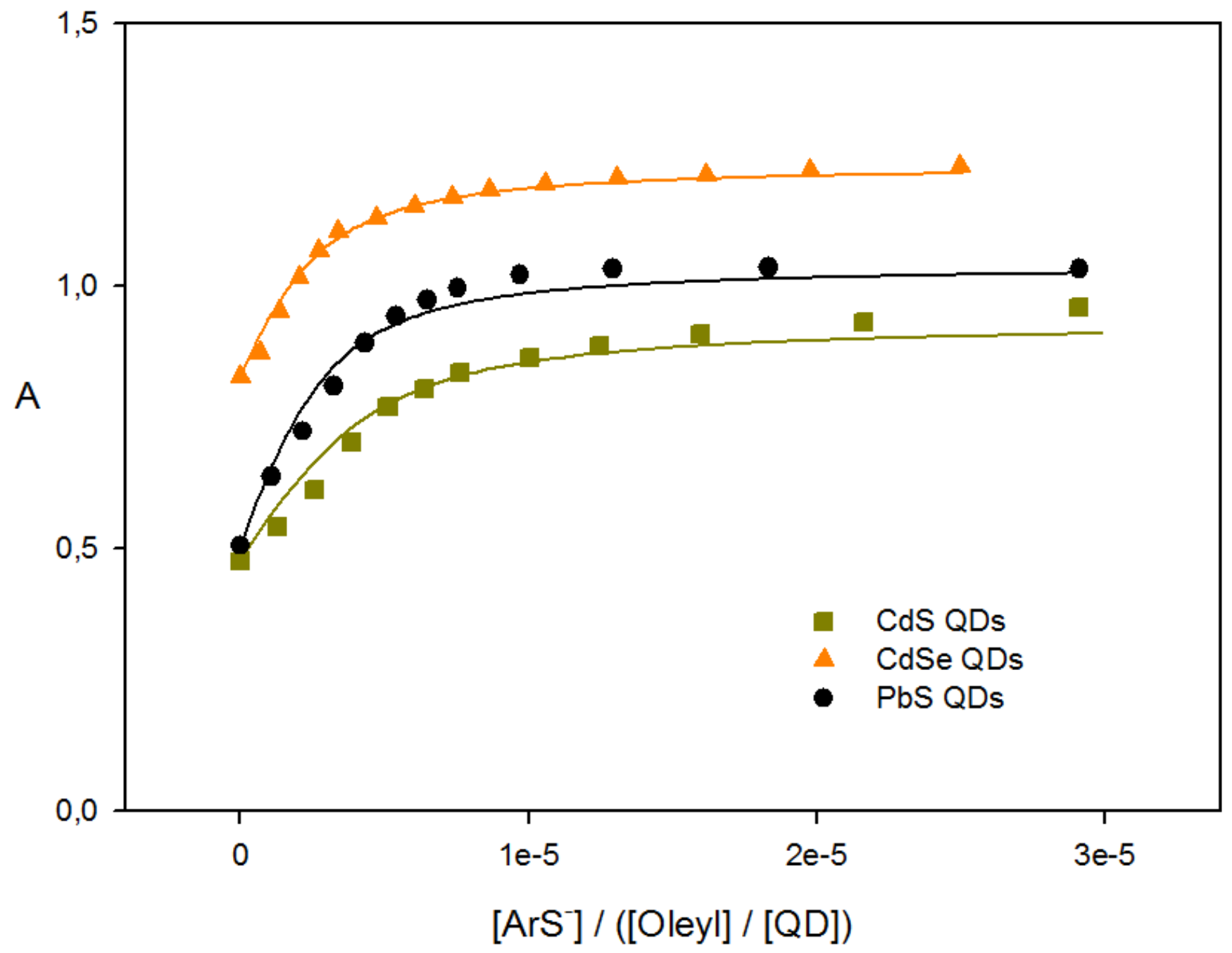

**Figure S7.** Plots of QD absorbance at wavelengths reported to correspond to bulk-like values, i.e. 310 nm for CdS, 340 nm for CdSe, and 400 nm for PbS, upon $ArS^{-}$ addition; lines represent fitting according to Supplementary equations S1-5.

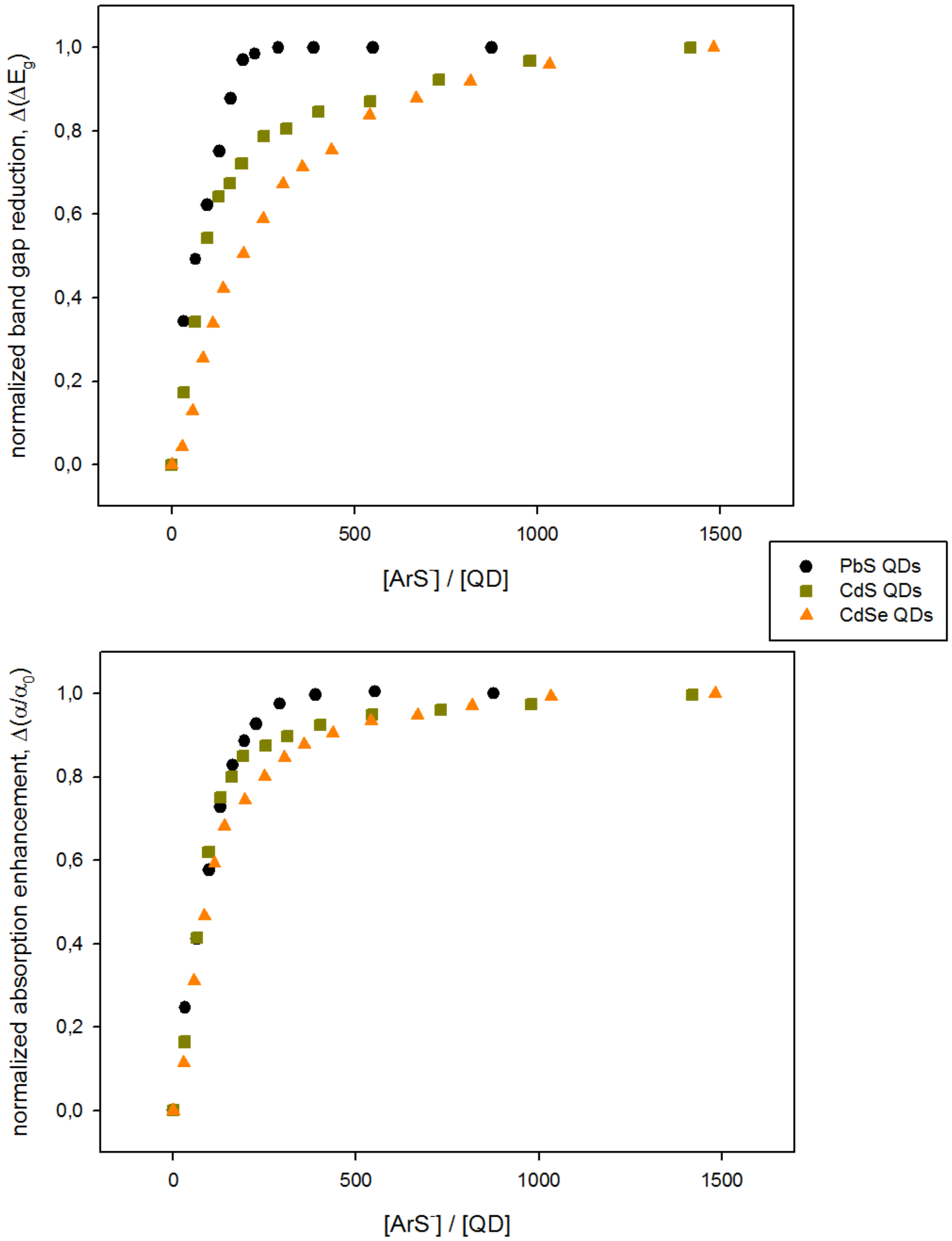

**Figure S8.** (top panel) Normalized plots of the $ArS^{-}$-induced QD optical band gap reduciton ($\Delta E_g$, as the difference between first exciton peak energy after and before ligand exchange). (bottom panel) Normalized plots of the $ArS^{-}$-induced QD broadband optical absorption enhancement ($\alpha/\alpha_0$, as the ratio of energy integrated absorption coefficients after and before ligand exchange).

In order to broaden the significance of this experimental observation, ArS¯ was added also to wurzite Cd chalcogenide QDs: minor spectral changes are observed compared to those induced on zinc-blende Cd chalcogenide QDs (Figures S9 and S10); whether this is due to the ineffective replacement of octadecylphosphonate ligands coming from the synthetic procedure or to the different crystal lattice cannot be straightforwardly asserted. However, tighter binding of phosphonates than carboxylates at CdSe QD surface was already demonstrated,[S21] whereas the ArS¯-induced spectral changes on wurzite CdSe QDs are here shown as rather similar to those induced by sterically encumbered ligands, SE-ArS¯, on zinc-blende CdSe QDs (Figure S9), thus suggesting the relevance of inefficient exchange of octadecylphosphonate ligands with ArS¯ in wurzite CdS and CdSe QDs.

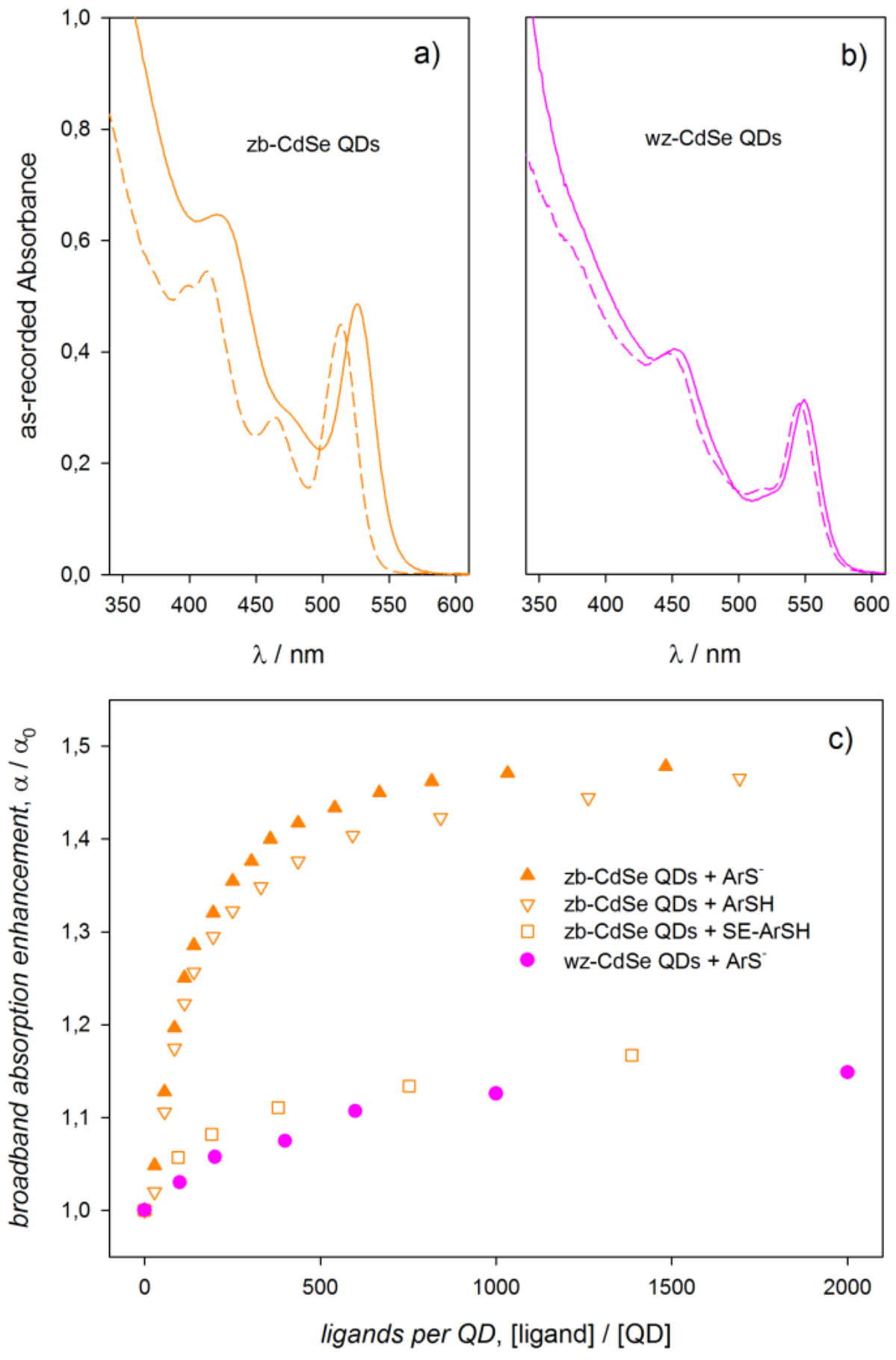

**Figure S9.** a) As-recorded absorption spectra of as-synthesized zinc-blende CdSe QDs (dashed line) and upon addition of ArS¯ (solid line). b) As-recorded absorption spectra of as-synthesized wurzite CdSe QDs (dashed line) and upon addition of ArS¯ (solid line). c) Plots of the ligand-induced CdSe QD broadband optical absorption enhancement ($\alpha/\alpha_0$) as indicated in the legend.

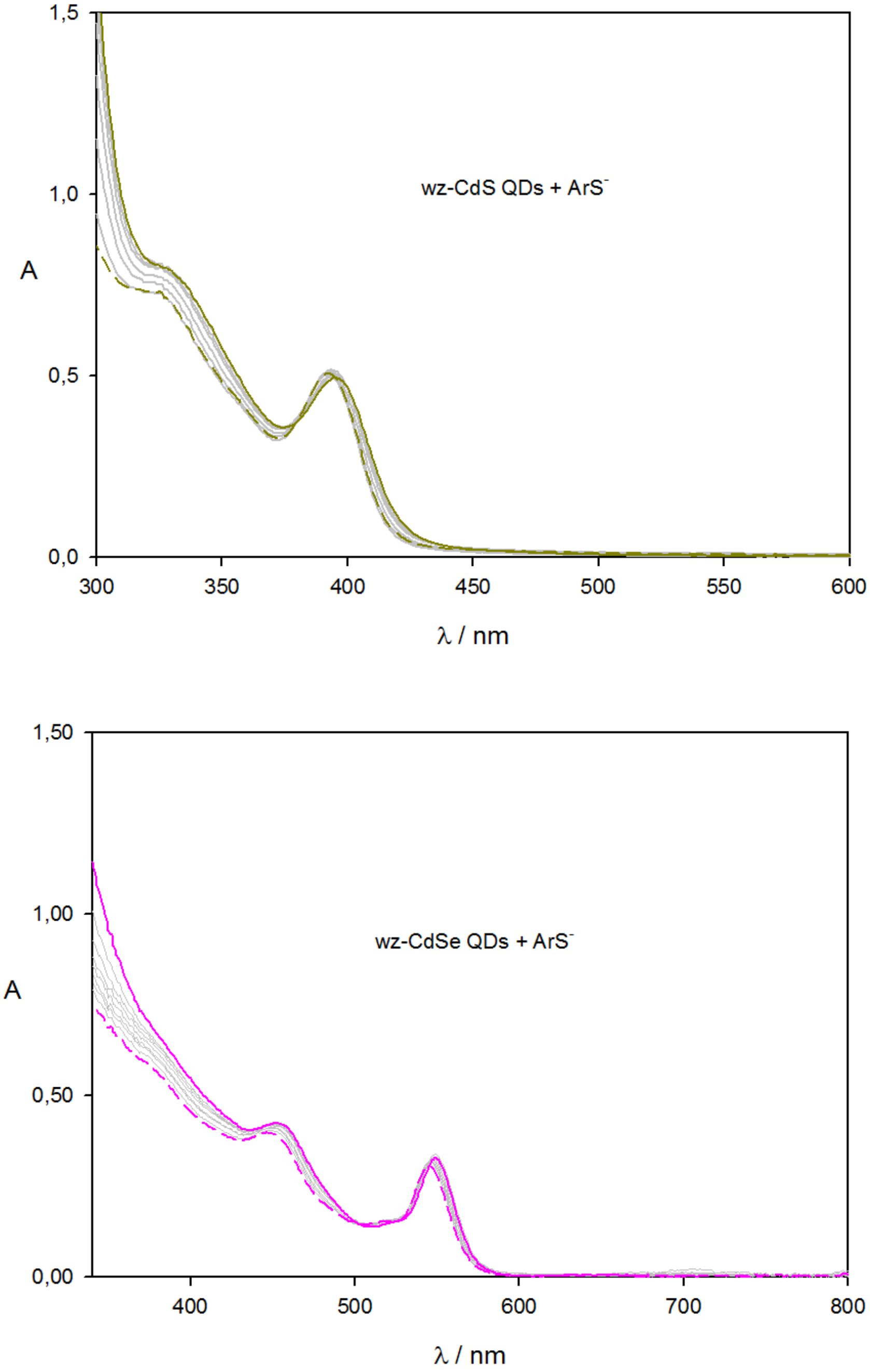

**Figure S10.** Absorbance spectra of wurzite (top panel) CdS and (bottom panel) CdSe QDs (dashed lines) upon addition of p-methylbenzenethiolate.

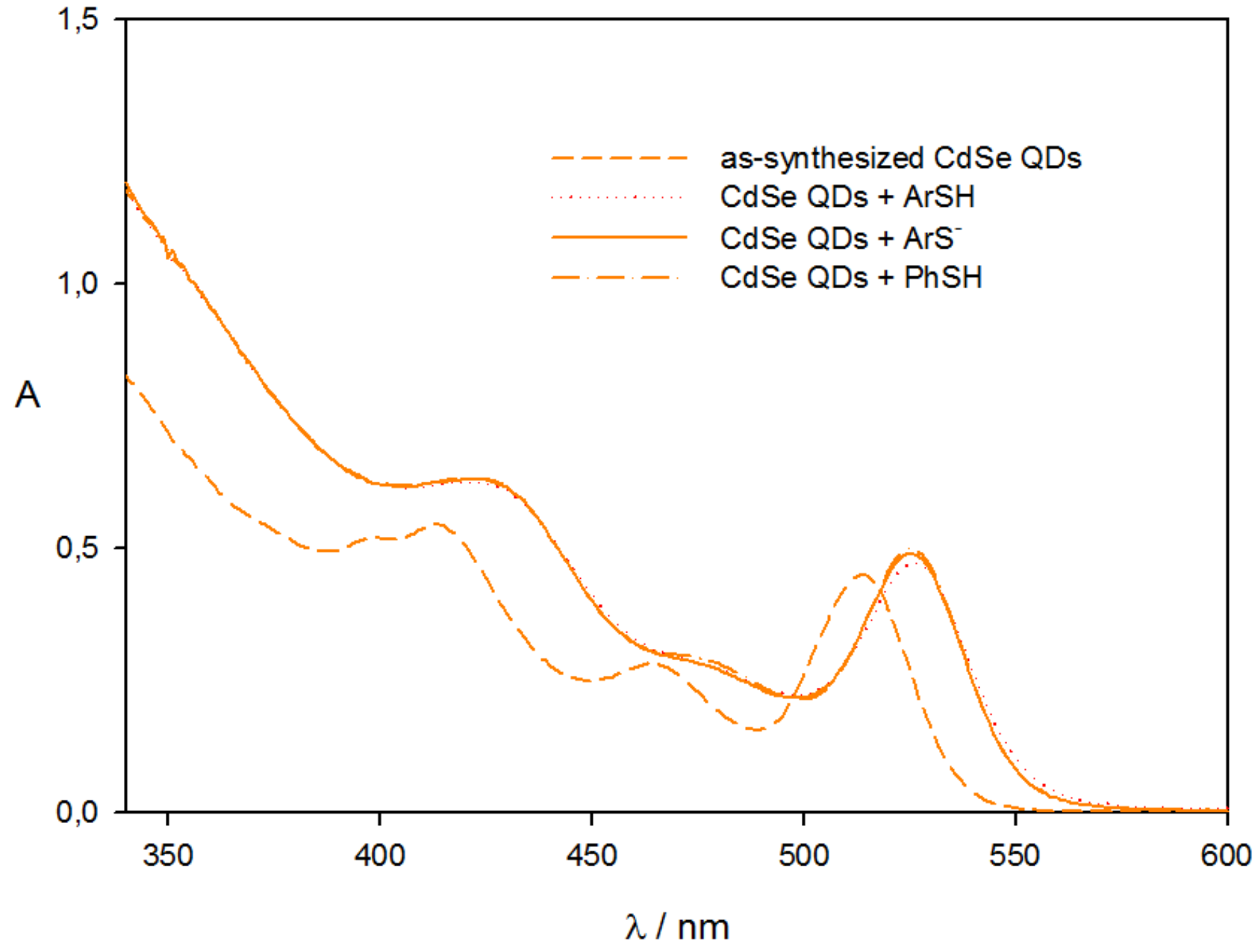

**Figure S11.** Absorbance spectra of zinc-blende CdSe QDs (dashed line) and upon addition of $ArS^{-}$, ArSH, and PhSH.

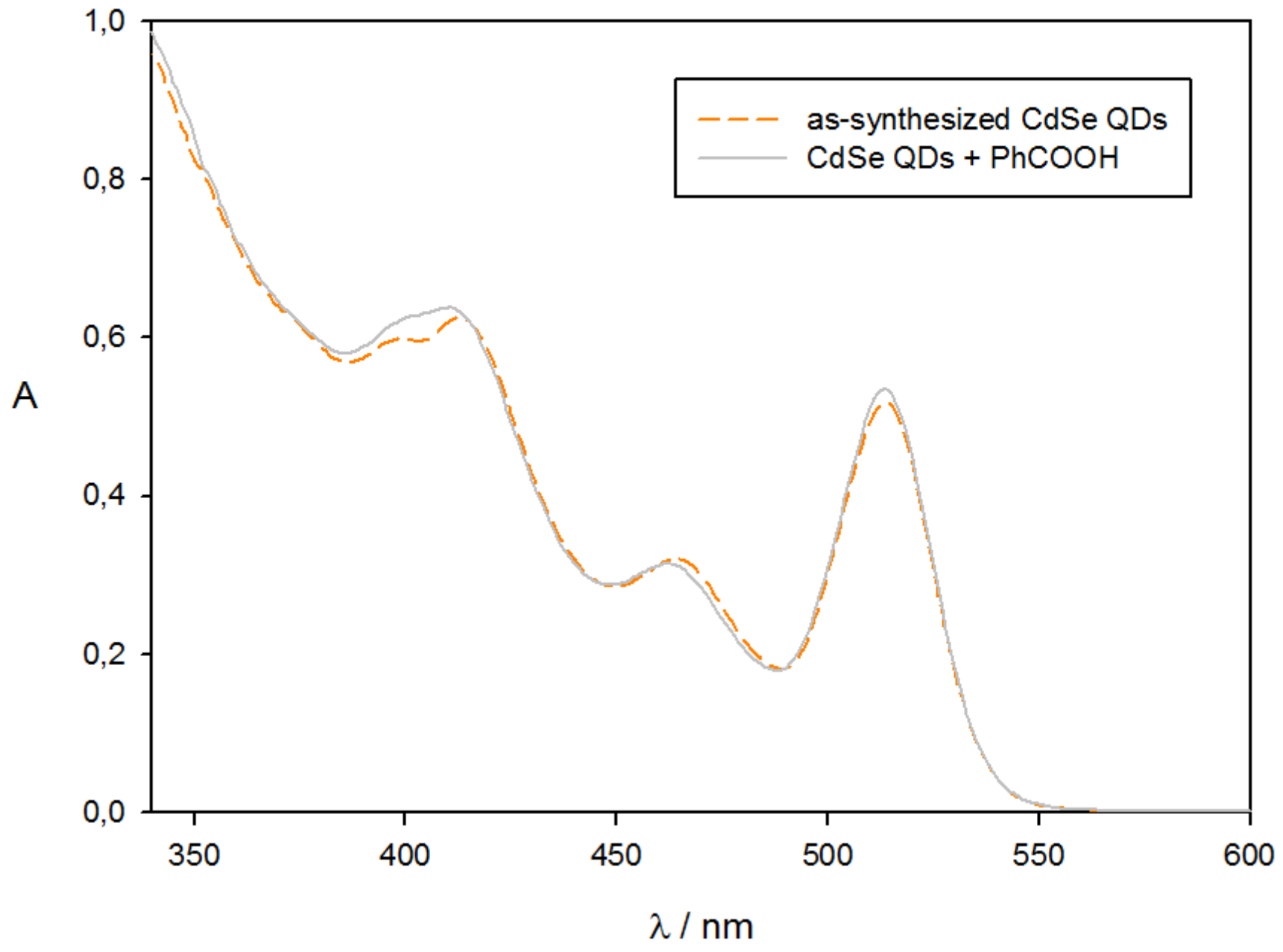

**Figure S12.** Absorbance spectra of zinc-blende CdSe QDs (dashed line) and upon addition of PhCOOH.

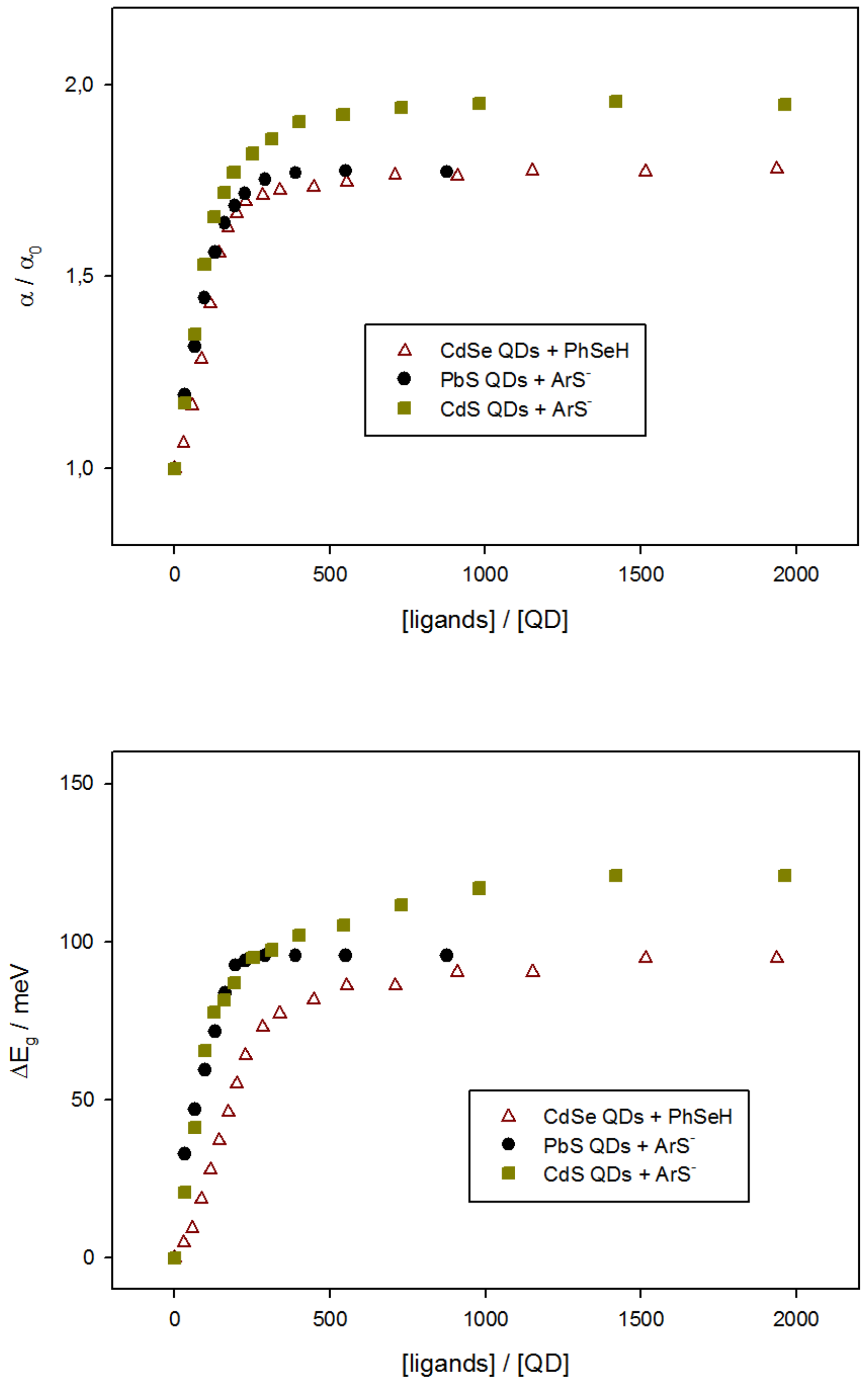

**Figure S13.** Plots of (top panel) broadband optical absorption enhancement and of (bottom panel) optical band gap reduction of CdS upon $ArS^{-}$ addition, CdSe upon PhSeH addition, and PbS upon $ArS^{-}$ addition.

The core/shell dielectric mismatch may induce surface polarization affecting exciton energy and was evaluated as follows. Upon using a two-state nearly free-electron model within the framework of effective mass approximation,[S22] accounting for the energy of the first exciton transition in QDs, such polarization effect can be described under the assumption that bulk transport gap, kinetic energies of the electron and the hole due to quantum confinement, and the electron/hole electrostatic interaction are related only to the inorganic core and are therefore unaffected by ligand exchange; the observed bathochromic shift, $\Delta E_g$, thus coincides with the difference in the polarization term upon ligand exchange, $\Delta\delta$, according to:[S23-25]

$$\Delta E_g \equiv \Delta\delta \tag{S6}$$

with,

$$\delta = \frac{\pi e^2}{\varepsilon_{QD}\varepsilon_0 d_{QD}} \sum_{l=1}^{\infty} \frac{d_{QD}}{2}^{2l+1} A_l \int_0^1 [j_0(\pi x)]^2 x^{2l+2} dx \tag{S7}$$

where $\varepsilon_0$ is the vacuum permittivity and $\varepsilon_{QD}$ is high-frequency[S26] dielectric constant of the core, $d_{QD}$ is the QD diameter, $e$ is the fundamental charge for the electron, $j_0(x)$ is the zero-order spherical Bessel function, and the term $A_l$ is given by:

$$A_l = \frac{l+1}{\frac{d_{QD}}{2}^{2l+1}} \frac{\frac{d_{QD}}{2}^{2l+1}(\varepsilon_\ell - \varepsilon_s)[\varepsilon_c + l(\varepsilon_c + \varepsilon_\ell)] + \left(\frac{d_{QD}}{2} + L_\ell\right)^{2l+1}(\varepsilon_c - \varepsilon_\ell)[\varepsilon_s + l(\varepsilon_\ell + \varepsilon_s)]}{\frac{d_{QD}}{2}^{2l+1}(\varepsilon_c - \varepsilon_\ell)[(\varepsilon_{lig} - \varepsilon_{solv})l(l+1)] + \left(\frac{d_{QD}}{2} + L_\ell\right)^{2l+1}[\varepsilon_\ell + l(\varepsilon_c + \varepsilon_\ell)][\varepsilon_s + l(\varepsilon_\ell + \varepsilon_s)]} \tag{S8}$$

in which the dielectric constants for the ligand shell, $\varepsilon_\ell$, are assumed as the square of the oleic acid and benzenethiol ligand refractive indexes, despite the severe approximation of attributing a dielectric constant to a sort of monolayer at the QD surface; the solvent high-frequency dielectric constant, $\varepsilon_s$, is taken as the square of the solvent refractive index;[S27] the ligand lengths, $L_\ell$, are assumed as 1.8 nm for native oleate and 0.6 nm for replacing benzenethiolate. The optical band gap reduction expected from dielectric arguments is lower than a couple of meV for the size range here considered, as shown in Figure 4a and S14 (for PbS QDs).

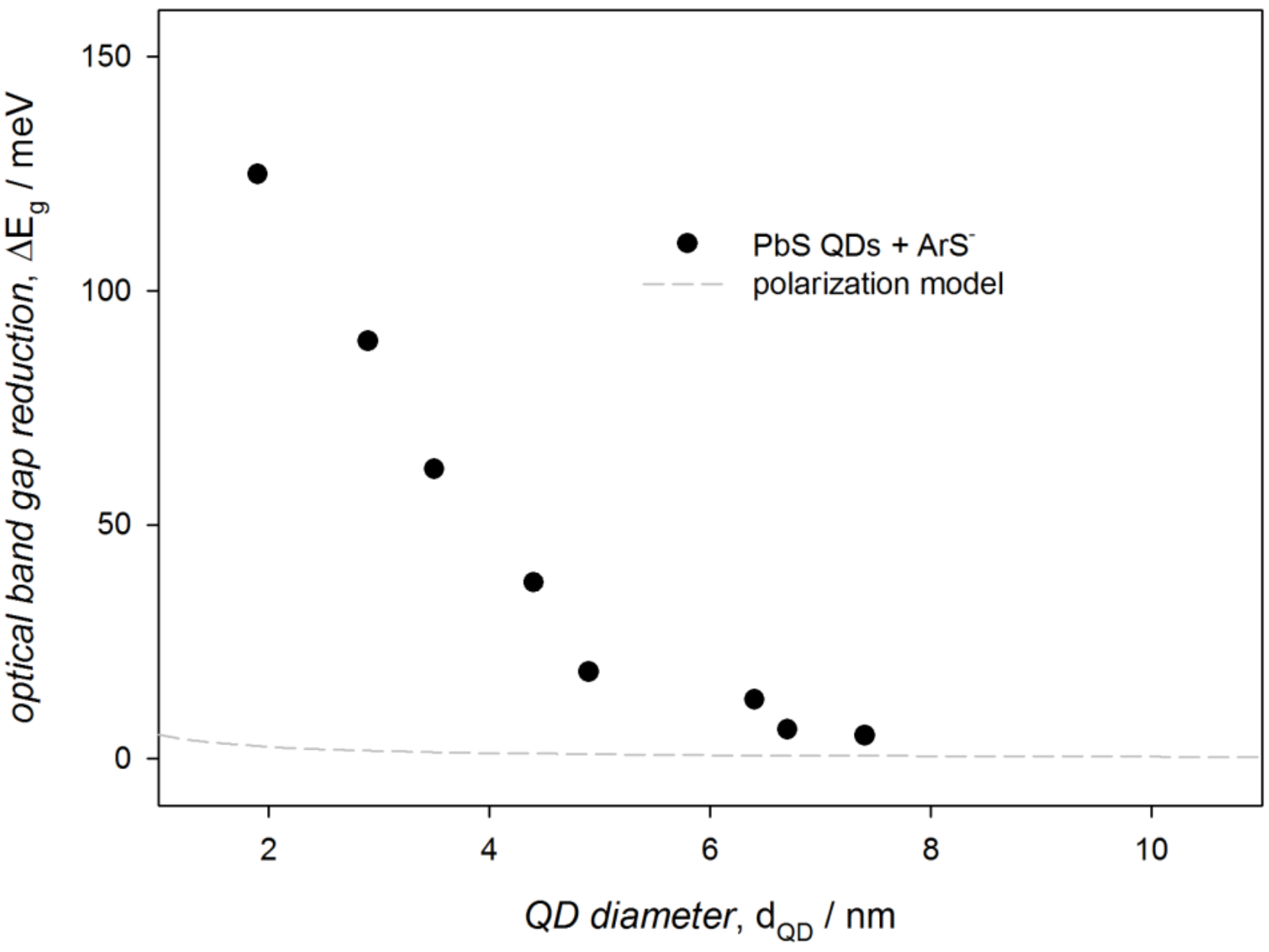

**Figure S14.** Plot of the first exciton peak red shift observed upon addition of $ArS^-$ to solutions of colloidal PbS QDs. Dashed line is the polarization effect predicted according to equations S6-8 upon exchanging native oleate ligands for $ArS^-$ on PbS QDs.

The height of the potential barrier at the core interface with ligands may affect quantum confinement thus changing the QD first exciton energy and was evaluated as follows. Upon using a double spherical finite potential well model[S28-S30] within the framework of effective mass approximation[S22] accounting for the energy of the first exciton transition in QDs, the optical band gap reduction was related to the different kinetic energies of one of the charge carriers (the hole, presumably) under the assumption that bulk transport gap, kinetic energies of the other charge carrier (the electron, consequently) due to quantum confinement, and the electron/hole electrostatic interaction are unaffected by ligand exchange:

$$\Delta E_g \equiv \Delta E_h \tag{S9}$$

For the charge carrier under the action of a force:

$$\left[-\frac{\hbar^2}{2}\nabla\frac{1}{m}\nabla + V(\vec{r})\right]\psi(\vec{r}) = E_h\psi(\vec{r}) \tag{S10}$$

For a symmetric, stepwise potential:

$$V(\vec{r}) = V(r) = \begin{cases} V_c\,, & 0 \le r \le \dfrac{d_{QD}}{2} \\ V_\ell\,, & \dfrac{d_{QD}}{2} < r \le \dfrac{d_{QD}}{2} + L_\ell \\ V_s\,, & \dfrac{d_{QD}}{2} + L_\ell < r \end{cases} \tag{S11}$$

With $V_c = 0$, $V_\ell = 0$ and 2 eV for oleate and benzenethiolate, respectively, and $V_s = 5$ eV.

And the radial and spherical coordinates can be separated:

$$\psi(r,\theta,\phi) = \frac{1}{\sqrt{4\pi}} R(r) \tag{S12}$$

For the ground state (n = 1, l = m = 0).

$$R(r) = \begin{cases} A_c j(kr) + B_c n(kr), & 0 \le r \le \frac{d_{QD}}{2} \\ A_\ell h^{(+)}(ikr) + B_\ell h^{(-)}(ikr), & \frac{d_{QD}}{2} < r \le \frac{d_{QD}}{2} + L_\ell \\ A_s h^{(+)}(ikr) + B_s h^{(-)}(ikr), & \frac{d_{QD}}{2} + L_\ell < r \end{cases} \tag{S13}$$

With *j, n, h* denote spherical Bessel, Neumann, and Hankel functions, respectively, and

$$k = \sqrt{\frac{2m|E_h - V|}{\hbar^2}} \tag{S14}$$

In the core, ligand, solvent regions, in which potential barrier height is defined above and m are the effective masses assumed as $m_e$ for both ligand and solvent regions.

The solution must be continuous and derivable, regular for r = 0 and vanish at infinity, thus yielding a system of four linear equations for the four coefficients with non-trivial solutions if its determinant is 0; normalization permits to determine such coefficient values.

The optical band gap reduction expected from exciton delocalization for QDs with 3.0 nm diameter bearing oleate and benzenethiolate ligands is shown in Figures 4b and S15 (for PbS QDs).

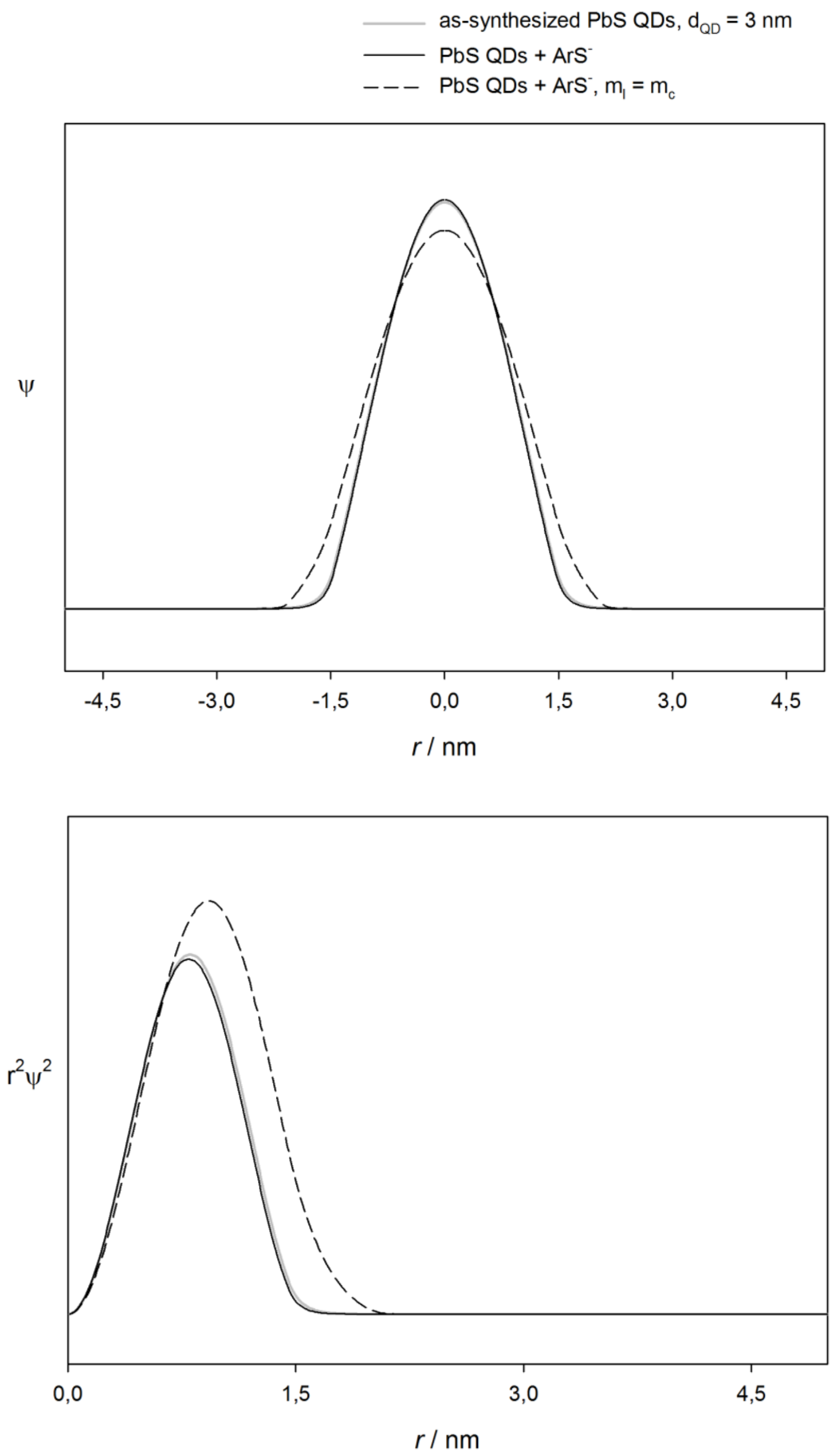

**Figure S15.** (top) wavefunctions in a 3.0 nm diameter PbS QD with oleate ligands ($V_l$ = 2 eV, grey solid line), with benzenethiolate ligands ($V_l$ = 0 eV, black solid line); and with benzenethiolate ligands ($m_l$ = $m_c$ = 0.085, black dashed line). (bottom) radial wavefunction distribution as in the top panel.

The experimentally observed optical absorption enhancement induced by surface ligands highlights the limitations of formalisms mutuated from classic electromagnetism in describing the linear optical absorption of colloidal semiconductor nanocrystals. Indeed, the optical absorption of dilute dispersions of small colloidal crystals (where dilute stands for non-interacting particles –the volume fraction is much lower than one– and small indicates that the particles are much smaller than the wavelength of the incident light) surrounded by a dielectric medium (with much lower refractive index than the particle itself) is commonly described using formalisms derived from Maxwell Garnett effective medium theory.[S31] In this framework, the intrinsic absorption coefficient at energies sufficiently far from the band gap matches, up to a multiplicative constant, the bulk value rescaled for the Lorentz local field factor that relates the external, applied electric field and the electric field within the inorganic core, which is affected by the surroundings (the ligand shell and the solvent; details are given in the Supplementary Information):[S32]

$$\mu_{\lambda} \propto \alpha_{\lambda} \left| f_{LF} \right|^2 \tag{S15}$$

where $\alpha_{\lambda}$ is the thickness-independent absorption coefficient of homogeneous, bulk ME and $f_{LF}$ represents the Lorentz local field factor that relates the external, applied electric field and the electric field within the QD, accounting for the dielectric effect of the surroundings (the ligand shell and the solvent) on as-synthesized ME QDs, ${f_{LF}}^0$, and on ligand-exchanged ME QDs, $f_{LF}$ . Such factor can be used to evaluate the intrinsic absorption coefficients at wavelengths with expected bulk behavior with an expression derived for core/shell QDs,[S33] albeit the severe approximation of assuming a dielectric constant for a sort of organic ligand monolayer at the inorganic core surface:

$$\left| f_{LF} \right|^2 = \left| \frac{9\varepsilon_{lig}\varepsilon_{solv}}{a\varepsilon_{lig} + 2b\varepsilon_{solv}} \right|^2 \tag{S16}$$

with,

$$a = \varepsilon_{core}\left(3 - 2\frac{\left(L_{lig} + \frac{d_{QD}}{2}\right)^3 - \left(\frac{d_{QD}}{2}\right)^3}{\left(L_{lig} + \frac{d_{QD}}{2}\right)^3}\right) + 2\varepsilon_{lig}\frac{\left(L_{lig} + \frac{d_{QD}}{2}\right)^3 - \left(\frac{d_{QD}}{2}\right)^3}{\left(L_{lig} + \frac{d_{QD}}{2}\right)^3} \quad \text{(S17)}$$

$$b = \varepsilon_{core}\frac{\left(L_{lig} + \frac{d_{QD}}{2}\right)^3 - \left(\frac{d_{QD}}{2}\right)^3}{\left(L_{lig} + \frac{d_{QD}}{2}\right)^3} + \varepsilon_{lig}\left(3 - \frac{\left(L_{lig} + \frac{d_{QD}}{2}\right)^3 - \left(\frac{d_{QD}}{2}\right)^3}{\left(L_{lig} + \frac{d_{QD}}{2}\right)^3}\right) \quad \text{(S18)}$$

where the high-frequency dielectric constant value is assumed as the dielectric constant for the ME core, $\varepsilon_{core}$; the square of the oleate and p-methylbenzenethiolate ligand refractive indexes and of dichloromethane (tetrachloroethylene) are assumed as the dielectric constants for the ligand shell and the solvent, $\varepsilon_{lig}$ and $\varepsilon_{solv}$, respectively; 1.8 nm for oleate and 0.6 nm for p-methylbenzenethiolate are assumed as the ligand lengths, $L_{lig}$ .

The optical absorption enhancement, $\mu_{\lambda} / \mu_{\lambda}^{0}$, ascribable to changes in the internal electric field expected upon exchanging oleate ligands for p-methylbenzenethiolate species is small, when calculated as the local field factor ratio for ligand-exchanged and as-synthesized PbS QDs, $\left|f_{LF}\right|^2 / \left|f_{LF}^{0}\right|^2$ (see Figures 4c ans S16). The ligand-induced enhancement of QD optical absorption experimentally observed upon exchanging oleates for p-methylbenzenethiolates is instead much larger than expected by mere dielectric confinement, as reported in Figure S15 in the case of PbS QDs.

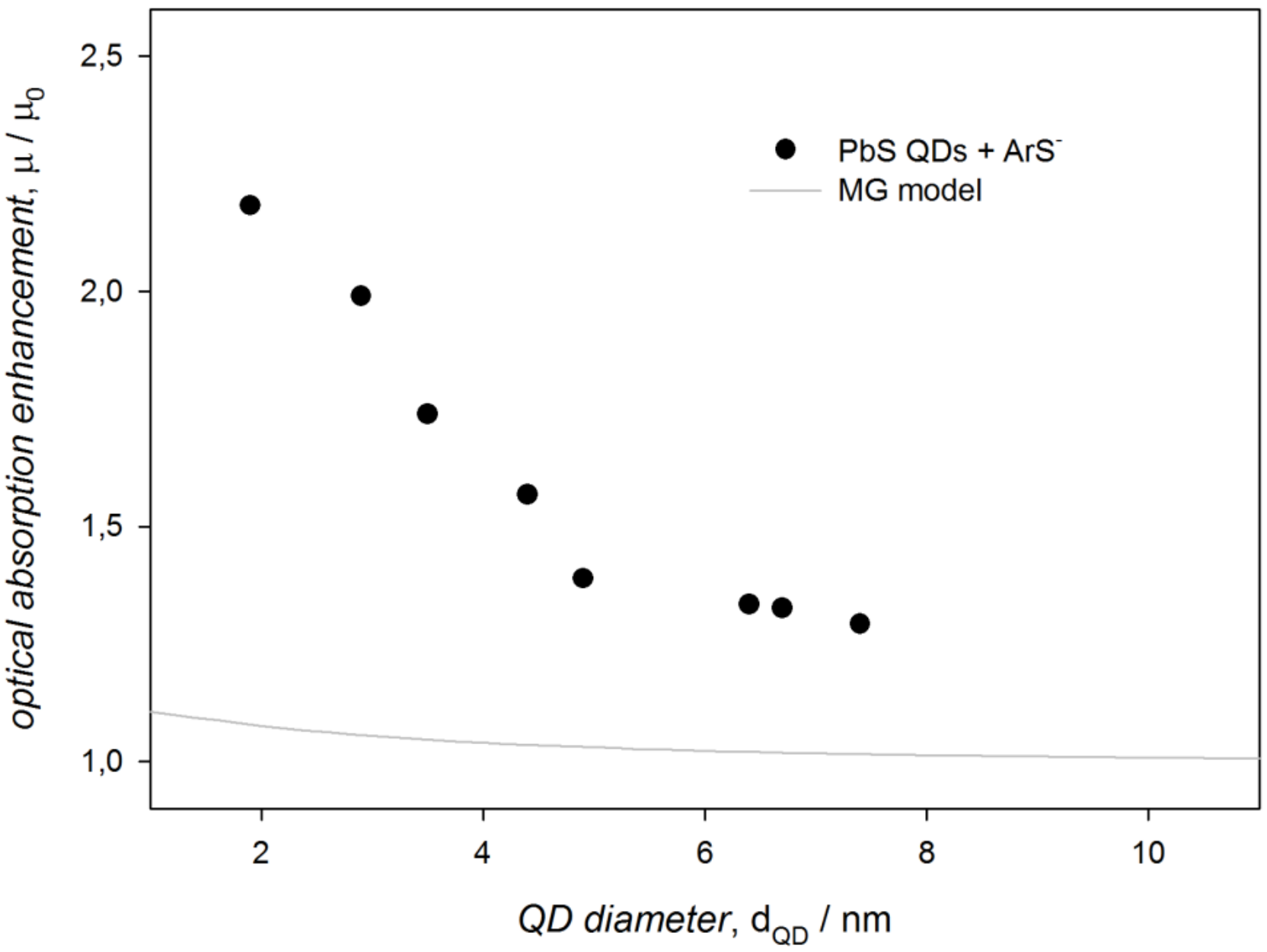

**Figure S16.** Experimentally determined (symbols) intrinsic absorption coefficient ratio at 400 nm of ligand-exchanged PbS QDs, $\mu_{400}$, and as-synthesized QDs, $\mu_{400}^{0}$, as function of QD size, $d_{QD}$; dashed line represents the local field factor ratio between ligand-exchanged and as-synthesized PbS QDs as function of $d_{QD}$, according to equations S15-18.

## Supplementary References.